\documentstyle[aps,prl,epsfig,amsfonts]{revtex}
\makeatletter
\newlength\abovecaptionskip \newlength\belowcaptionskip
\def\@makecaption#1#2{%
 \vskip\abovecaptionskip \sbox\@tempboxa{#1: #2}%
 \ifdim \wd\@tempboxa >\hsize #1: #2\par \else \global \@minipagefalse
 \hb@xt@\hsize{\hfil\box\@tempboxa\hfil}%
 \fi \vskip\belowcaptionskip} \makeatother

\newcommand{\fr}{\frac}
\newcommand{\lan}{\langle}
\newcommand{\ran}{\rangle}

\newcommand{\eps}{\varepsilon}

\title{Level statistics and eigenfunctions of pseudointegrable systems:
dependence on energy and genus number}
\author{Yuriy Hlushchuk and Stefanie Russ\\
Institut f\"ur Theoretische Physik III, Universit\"at Giessen, \\
D-35392 Giessen, Germany}
\date{\today}
\begin{document}            

\draft
\maketitle
\begin{abstract} 
We study the level statistics (second half moment $I_0$ and rigidity 
$\Delta_3$) and the eigenfunctions of pseudointegrable systems 
with rough boundaries of different genus numbers $g$.
We find that the levels form energy intervals with a characteristic 
behavior of the level statistics and the eigenfunctions in each interval.
At low enough energies, the boundary roughness is not resolved and 
accordingly, the eigenfunctions are quite regular functions and the 
level statistics shows Poisson-like behavior. 
At higher energies, the level statistics of most systems moves from 
Poisson-like towards Wigner-like behavior with increasing $g$. 
Investigating the wavefunctions, we find many chaotic functions that 
can be described as a random superposition of regular wavefunctions.
The amplitude distribution $P(\psi)$ of these chaotic functions was 
found to be Gaussian with the typical value of the localization volume 
$V_{\rm{loc}}\approx 0.33$. 
For systems with periodic boundaries we find several additional energy 
regimes, where $I_0$ is relatively close to the Poisson-limit. 
In these regimes, the eigenfunctions are either regular or localized 
functions, where $P(\psi)$ is close to the distribution of a sine or 
cosine function in the first case and strongly peaked in the second case.
Also an interesting intermediate case between chaotic and localized 
eigenfunctions appears.
\end{abstract}
]

\section{Introduction}

Quantum billiards are quite simple models for 
many practical applications in solid state and nuclear physics,
as e.g. quantum dots, microdisk lasers and electron 
transport in microstructures. 
An important means for the study of quantum billiards is the statistics of 
the quantum mechanical energy levels of a given system, where 
the distance distribution $p(s)$ of the normalized distances 
$s_\alpha=(E_{\alpha+1}-E_\alpha)/\lan s \ran$ between two consecutive energy 
levels $E_{\alpha+1}$ and $E_\alpha$ with the mean distance $\lan s \ran$ 
has the following two limiting cases. 
(i) The Poisson distribution, $p_P(s) = \exp[-s]$, which is the distance 
distribution of uncorrelated numbers $E_\alpha$ and (ii) the Wigner distribution 
$p_W(s) = \pi s/(2\lan s \ran^2)\exp[-\pi s^2/(4\lan s \ran^2)]$.
An example for the case (i) are the energy levels of a single symmetry 
group of an electron in a 2D potential well in the shape of a square, 
rectangle or circle (integrable systems). As a second example, localized 
states in disordered systems tend to the Poisson distribution with 
increasing system size, which is largely used in solid state physics to 
distinguish between localized and extended states in disordered systems 
\cite{guhr}.
An example for the case (ii) are the energy levels of chaotic billiards,
as e.g. the stadium or the Sinai billiard.

In close analogy is the behavior of a classical particle in a billiard, which 
undergoes elastic reflections at the walls. 
It also has two limiting cases, depending on the billiard geometry. 
If the billiard is chaotic, the motion of the particle is ergodically 
extended over the whole energy surface in phase space. 
Two particles whose trajectories are very close at the beginning, 
diverge exponentially from each other. If the system is integrable on
the other hand, the motion of the particle is restricted to a
two-dimensional torus in phase space and neighboring trajectories 
diverge only linearly from each other. 

There are, however, several classes of intermediate systems between the 
two limiting cases, as e.g. polygonal pseudointegrable systems 
\cite{pseudo1,pseudo2,pseudo3}, on which we focuse in this paper,
or systems with a pointlike scatterer (''Seba-Billiards'') \cite{seba}. 
Like in integrable systems, the motion of a classical particle in a 
pseudointegrable system is restricted to a two-dimensional surface in phase 
space. 
However, these surfaces do not have the shapes of tori but are more complicated 
objects with more than one hole. They are called ''multihandeled spheres''.
Examples for pseudointegrable systems are polygons with only rational 
angles $n_i\pi/m_i$, with $n_i, m_i\in \mathbb{N}$ and at least one $n_i>1$. 
They are described by their genus number 
\begin{equation}\label{genus}
g = 1 + \frac{M}{2} \sum_{i=1}^{J}\frac{n_i-1}{m_i},
\end{equation}
which is equal to the number of holes in the multihandeled sphere in
phase space. 
Here, $J$ is the number of angles and $M$ is the least common multiple of the 
$m_i$. The reason that those systems are not completely integrable is their 
property of beam splitting. 
At some points in their geometry, neighboring trajectories of particles can be 
split into two opposite directions (see Fig.~\ref{fig1}). 

Several works \cite{cc,shudoshim,shudoetal,steffi,BiswasJain,BiswasSinha} found the 
distance distribution of pseudointegrable systems intermediate between 
the Poisson- and the Wigner distribution. 
Ref. \cite{cc} presented a numerical investigation of pseudointegrable 
billiards of small genus numbers, where the boundary was approached by 
a step function, which was arranged 
along the curved shape of the (chaotic) Sinai billiard.
With decreasing size and increasing number of the steps, 
the level statistics of this system approached the Wigner statistics. 
However, as it was pointed out in this work, the analysis was performed 
at very small energies, where the wavelengths were larger than the step 
sizes. 
So, it was assumed that the Wigner-like level statistics was not an
inherent feature of the pseudointegrable shapes, but came from the underlying
assymptotic shape of the Sinai billiard. 
Pseudointegrable systems of small genus numbers $g=2$ and $3$ were investigated
numerically \cite{shudoshim,shudoetal} and experimentally
\cite{shudoetal}, showing an intermediate level statistics that clearly deviated
from the Wigner distribution.
In \cite{steffi}, numerical simulations on systems
with increasing genus numbers up to $g\approx 1000$ indicated that 
for not too small energies, the distance distribution changes 
systematically from Poisson-like towards Wigner-like behavior with
increasing $g$. 
Also for Seba billiards, it was shown by general arguments and numerical
calculations that the level statistics is intermediate and close to 
Wigner-like behavior for small level distances \cite{seba,shig94,weaver95,chsh96}. 
The Wigner-like behavior is increased
with the number and the coupling strength of the pointlike scatterers in 
the system \cite{shig94,chsh96}.

The level statistics seems to be connected to the properties of 
the eigenfunctions. For example, in a billiard with $g=2$ 
it was found that there are regular and irregular 
eigenfunctions coexisting \cite{BiswasJain}. In the Husimi representation,
the eigenfunctions of systems with small $g$ show signatures of
pseudointegrability, whereas those of systems with large $g$ tend to be 
irregular \cite{BiswasSinha}.

In this paper, we want to investigate the energy dependence of the level 
statistics of a special class of pseudointegrable systems with high genus
numbers. 
At low energies, it is known that one can observe a level statistics
that deviates from the high-energy limit, as shown in Ref.~\cite{cc} for
pseudointegrable and in Refs.~\cite{shudoshim,Borgonovi,Frahm1,Frahm2}
for chaotic billiards. 
Here we find that for our systems, 
there can be many energy windows, where the level statistics 
is comparatively close to Poisson statistics, and other energy intervals, 
where the behavior is close to Wigner statistics.
We show that this behavior is correlated to the properties of the 
eigenfunctions and that there exist several characteristic types of 
such intervals, even at energy values, where the boundary
roughness is resolved. 

The paper is organized as follows: In section II, we introduce the special 
pseudointegrable geometries that we consider.
In section III, we show the results of the level statistics of the
eigenvalues for several systems of different values of $g$.
In section IV, we investigate the eigenfunctions $\Psi^{(\alpha)}$ in these 
energy intervals and introduce several quantitative measures, i.e., 
the localization volume $V_{\rm{loc}}^{(\alpha)}$, the amplitude distribution 
$P(\psi^{(\alpha)})$ and the behavior of the eigenfunctions in $n,m$-space. 
The functions $\psi$ and $\Psi$ differ in their normalization, $\psi=\sqrt{A}\Psi$,
with the area $A$ of the system.
Finally, in section V, we apply those measures to many eigenfunctions in the 
different energy intervals.
We find that in intervals where the energy-levels behave Poisson-like, 
the eigenfunctions are either localized or regular.
Energy intervals with level statistics close to Wigner-like behavior, 
on the other hand, contain eigenfunctions that are random superpositions 
of plane waves. 
Also an interesting mixed case appears, where the superposition of plane 
waves leads to weak localization.

\section{Systems and Calculations}

We consider a membrane of the sound velocity $c$ that lies in the 
$xy$-plane and vibrates in the $z$-direction.
When the restoring forces are considered as scalar, the vibrations of 
this membrane are described by the Helmholtz equation
\begin{equation} \label{basic}
\Delta\Psi^{(\alpha)}(x,y) =
-\fr{\omega_\alpha^2}{c^2}\Psi^{(\alpha)}(x,y),
\end{equation} 
with the $\alpha^{th}$ eigenfunction $\Psi^{(\alpha)}(x,y)$, 
the corresponding eigenvalue $\omega_\alpha^2$.
The boundary conditions can be of Dirichlet or of Neumann type, 
referring to a membrane that is kept fixed at the boundary or that 
can vibrate freely, respectively.
Equation (\ref{basic}) has the same form as the stationary Schr\"odinger 
equation with zero potential $V=0$ inside the system.
Therefore, under Dirichlet boundary conditions, which refer to an 
infinite potential on the boundary, it also describes an electron 
of mass $\mu$ in 
an infinite potential well. In this case, one has to replace 
$\omega_\alpha^2/c^2$ by $2\mu E_\alpha/\hbar^2$, with 
the energy eigenvalue $E_\alpha$.

For the numerical calculations, Eq.~(\ref{basic}) is discretized on a 
square lattice, which reduces the problem
to the diagonalization of a symmetric matrix, 
which is carried out here by the Lanczos algorithm \cite{Lanczos}, a numerical 
procedure to compute eigenvalues und -vectors of sparse $N\times N$-matrices 
by reducing them iteratively to a tridiagonal form, for which effective 
algorithms exist.
The eigenvalues $E_\alpha$ are calculated numerically under Dirichlet and Neumann 
boundary conditions and their spectra are analyzed by the means of level 
statistics. The $E_\alpha$ are dimensionless quantities, as we set $\hbar^2/(2\mu)
=1$ and $d=1$, where $d$ is the lattice constant of the discrete lattice.

As a model for our studies we chose rational billiards of the shape 
presented in Fig.~\ref{fig1} with different parameters, refering to
different numbers, widths, distances and heights of the ''teeth''.
The genus numbers of these geometries can be easily increased by increasing 
the number of teeth.
Two different angles occur, $\varphi=\pi/2$ and $\varphi=3\pi/2$.
Applying Eq.~(\ref{genus}) we find that $g=1+G_i$, where $G_i$ is the 
number of angles of values $\varphi=3\pi/2$. 
The systems have no symmetry axes and it is therefore not necessary to 
separate the calculated eigenvalues according to their symmetry groups 
(which would be technically difficult).
The considered parameters range from systems with only few teeth and 
therefore very small genus numbers up to systems of $g=101$. 

\section{Level Statistics}

We now analyze the eigenvalues by the means of level statistics. The 
energy levels $E_\alpha$ are normalized (''unfolded'') to new 
values $\eps_\alpha$, such
that their mean distance $\lan s\ran$ is equal to $1$.
Then, we calculate the following two quantities:

(i) From the nearest neighbor spacing distribution $p(s)$, it has 
become common to calculate the second half moments
\begin{equation}\label{i0}
I_0={1\over2}\left<s^2\right>={1\over2}\int_{0}^{\infty}s^{2}p(s)ds,
\end{equation}
which lie between the two limiting values $I_0^{\rm Wigner} 
\approx 0.637$ and $I_0^{\rm Poisson} = 1$  
(see Refs.~\cite{kantelhardt98,schweitzer99}). 
This enables us to decide, if the statistics is closer to Wigner or closer 
to Poisson by comparing just one number $I_0$ and is more comfortable in 
handling than $p(s)$ itself.  

(ii) Another measure, which turns out to be even more sensitive to $g$ is
the spectral rigidity $\Delta_3(L)$, where $L$ is the length 
of the considered energy interval \cite{dyson}. 
$\Delta_3(L)$ starts from the integrated density of states
$N(\eps)\equiv N(\eps_\alpha)=\sum_{n=1}^N \Theta(\eps_\alpha-\eps_n)$ 
of the unfolded energy levels $\eps_\alpha$, which is a staircase and can 
be approached by a straight line of slope one. 
$\Delta_3(L)$ is defined as the least square deviation,
\begin{equation}\label{delta3}  
\Delta_3(L) = \left\lan\rm{Min}_{r_1,r_2} \int_{\eps-L/2}^{\eps+L/2} 
[N(\eps)-r_1-r_2\eps]^2 d\eps  
\right\ran,  
\end{equation}  
where $\rm{Min}_{r_1,r_2}$ means that the parameters $r_1$ and $r_2$ are 
chosen such that the line $r_1+r_2\eps$ is the best fit of $N(\eps)$.
For the calculation of $\Delta_3(L)$ we use the technique derived in 
Ref.~\cite{bohigianno1975}.  
The limiting values are $\Delta_3(L)=L/15$ for integrable systems and 
$\Delta_3(L)=\ln(L)/\pi^2-0.07/\pi^2 + O(L^{-1})$ for the ensemble of 
Gaussian orthogonal matrices (GOE) \cite{methabuch}, which serves as a 
general accepted good limit for chaotic systems. 
This means that in the first case, $\Delta_3(L)$ increases linearly with 
$L$, and in the second case logarithmically. 
As discussed above, we expect intermediate behavior for pseudointegrable 
billiards.

In this paper, we will basically use $I_0$ as the easier of the two measures.
However, we first want to compare the behavior of $I_0$ and $\Delta_3$ 
for several systems, ranging from very small values of
$g=3$ until $g=49$ at higher energies. 
The parameters of these systems are shown in Tab.~\ref{table1}. 
All systems have roughly the same area and exactly the same widths and heights 
of the teeth, $a=h=8d$, where $d$ is the lattice constant. 

In Figs.~\ref{fig2}(a) and (c) we plotted the values of $I_0$ for 
Dirichlet and Neumann boundary conditions versus the energy. 
In Figs.~\ref{fig2}(b) and (d) we show the $\Delta_3(L)$-data of the fixed 
energy intervals $E (=\omega^2/c^2) \in[1.5,2.0]$ versus $L$. 
The energy interval is chosen such that the wavelength, 
$\lambda= 2\pi c/\omega$ is smaller than the widths of the 
teeth and therefore the boundary roughness is resolved.
The behavior of $I_0$ shows many fluctuations but its average value 
decreases systematically with growing $g$ towards the Wigner limit, 
which is shown as a solid line. 
Comparing Figs.~\ref{fig2}(a) and (c), we also see that the behavior for both
boundary conditions is quite similar. The only exception occurs at very small
energies, where $I_0$ lies considerably higher in the Dirichlet case. The reason
for this is that the condition $\psi=0$ exactly at the boundary prevents
the long-wavelengths eigenfunctions to penetrate into the small boundary teeths.
The same ''screening'' of the boundary roughness at small energies has already 
been observed for fractal drums \cite{damping}.
For $\Delta_3$ of pseudointegrable systems it was shown in \cite{biswas} 
by a semiclassical periodic orbit theory that it depends on several details, 
e.g. on the energy interval and on the area of the system. 
However, by keeping the billiard area, the energy interval and the height 
and width of the teeth fixed, we find also for $\Delta_3$ a smooth and 
systematic behavior, that depends on $g$.

The behavior of $\Delta_3(L)$ (see Figs.~\ref{fig2}(a) and (c)) corresponds 
to the one of $I_0$. 
Those systems with small genus numbers $g$ show high values of $I_0$ and 
accordingly values of $\Delta_3(L)$, which are close to 
$L/15$ (dotted line). 
Systems with high genus numbers $g$ on the other hand, show $I_0$-values 
close to $I_0^{\rm Wigner}$ and accordingly $\Delta_3$-curves, which are
also closer to the Wigner-limit (solid line). Similar to the case of Seba
billiards, they come very close to the Wigner distribution for small
$L$. $\Delta_3(L)$ turns out to be a more sensitive measure than $I_0$
in the case of large $g$-values. 
While the values of $I_0$ for systems with genus numbers $g>20$ lie 
already so close to the Wigner limit that a succession between them can 
hardly be recognized, we still observe clear differences between the 
individual curves of $\Delta_3$ in the case of larger level distances.
However, also the $\Delta_3$ data indicate a systematic change from 
Poisson-like towards Wigner-like behavior with increasing $g$.

In the following, we concentrate on $I_0$. 
We calculate $I_0$ for several selected systems with 
different heights, widths and numbers of the teeth in the energy range 
of $E\in [0,3]$, which corresponds to roughly $40000$ energy levels. 
The calculations were made in intervals of energy $\Delta E=0.05$, each 
of them 
containing about $600-800$ levels.
Here, we applied Neumann boundary conditions in all cases. 
The behavior of $I_0$ under Dirichlet boundary conditions is qualitatively 
similar. 

First, we discuss the case of random values of the parameters $a$, $b_x$, 
$b_y$ and $h$ (Tab.~\ref{table2}). 
For the first system $R_1$ the values were uniformly distributed between 
$4d$ and $15d$, for the second system $R_2$ between $4d$ and $10d$. 
The areas of the systems are again kept roughly constant.
In Fig.~\ref{fig3} (a), $I_0$ is plotted versus the energy $E$ 
for the systems $R_1$ (dashed lines) and $R_2$ (solid lines).
For most energy values, we find a roughly constant value of $I_0$ that lies 
close 
to the Wigner-value. 
Only for very low energy values, $E\in [0,0.2]$ (which corresponds to 
roughly $2100$ levels), $I_0$ shows deviations from the high energy behaviour, 
towards higher values. 
This is due to the finite resolution of the boundary roughness in the limit 
of long wavelengths, where the systems look more regular.
So, for a random structure of the boundary roughness, the high energy limit 
is reached quickly and no deviations of the $I_0$ values from the Wigner 
limit are seen beyond the first $2000$ states. 

In Fig.~\ref{fig3}(b-c), $I_0$ is plotted versus $E$ for two types of 
systems with periodic boundary roughness, whose parameters are presented 
in Tab.~\ref{table3}.  
In the first group (Fig.~\ref{fig3}(b)), all systems have the same widths 
$a=b_x=b_y=4d$ and numbers $N_x$ and $N_y$ of the ''teeth'' (and 
correspondingly the same genus number $g=87$). 
The heights of the teeth are different, $h=4d$ for system $B_1$ (solid 
lines), $8d$ for system $B_2$ (dotted lines) and $16d$ for system $B_3$ 
(dashed lines).
We first see that the narrow teeth account for a large low-energy regime 
with Poisson-like behaviour.
Additionaly, we observe sharp peaks of the $I_0$ values at several 
energies, which become more pronounced with increasing $h$ and are most 
probably due to the periodic structures of the geometry. 
We will discuss these peaks in the following sections.
The second group (cf. Fig.~\ref{fig3}(c)) consists of billiards with 
larger widths $a=b_x=b_y=8$. 
This group of billiards shows less and only small peaks in the 
$I_0$-values and the high energy regime is reached quite quickly.

Therefore we found two types of systems, where the high energy limit is 
reached quickly: geometries with broad teeth and geometries with a 
random distribution of the teeth widths. 
In systems with very narrow and regular teeth, on the other hand, energy 
windows appear, where the behavior of the system deviates significantly 
from the described behavior in the high-energy regime. 
In these intervals, the values of $I_0$ are considerably larger than the 
expected high-energy values. 
These energy windows are most interesting and we look at them in more 
detail now.

There are three different effects, which may lead to higher values of $I_0$.
First, at smaller energies (larger wavelengths), the teeth could not be 
sufficiently resolved and the states see only the rectangular main body of the system. 
This is the case for small energies, when half a wavelength is larger than the 
width $a$ and the eigenfunctions are not small enough to penetrate the teeth. 
Second, also at higher energy values, the periodic structure of the teeth
could allow for very regular functions, even if the boundary roughness is fully 
resolved. 
Also in this case, we expect a distribution close to Poisson.
Third, also localized states can be a reason for the spectrum to behave Poisson-like.
Therefore, we expect that $I_0$ is closely related to special system properties and 
should be reflected in the shape of the eigenfunctions.
In order to understand this, we now investigate the eigenfunctions in the
different energy windows.

\section{Eigenfunctions}

Some typical eigenfunctions of the system $B_3$ under Neumann boundary 
conditions are presented on Fig.~\ref{fig4}.
They are taken from the different energy regimes (labeled from $I$ to $IV$), 
as indicated in Fig.~\ref{fig4}(f). Due to technical 
restrictions of the Lanczos algorithm, we  could not calculate 
eigenfunctions of arbitrary high energies in large system sizes. 
The reason is that the density of states increases with the energy and 
the eigenfunctions become too close to each other and thus could not be 
seperated. At lower energies, we find the following characteristic shapes:

The eigenfunctions $\Psi^{(Ia)}$ and $\Psi^{(Ib)}$ of Fig.~\ref{fig4}(a) and (b) 
are taken from the energy regime $I$, 
where $I_0$ has the peak value of $I_0 = 0.942$, very close to the Poisson value. 
The functions look different.
While the function $\Psi^{(Ia)}$ on Fig.~\ref{fig4}(a) looks very regular and 
extended, function $\Psi^{(Ib)}$ on Fig.~\ref{fig4}(b) is  a rather localized 
(and regular) function, where non-zero amplitudes exist basically close to 
the boundary, i.e. inside the boundary teeth. 
Figure~\ref{fig4}(c) (regime $II$ on the Fig.~\ref{fig4}(f))
represents rather the case of a chaotic function, i.e. the amplitude looks 
very random. 
Accordingly, we find $I_0=0.688$ in this energy regime, closer to the 
Wigner value. 
The function $\Psi^{(III)}$ on Fig.~\ref{fig4}(d) (regime $III$ on the 
Fig.~\ref{fig4}(f)) looks again regular and $I_0=0.893$ is again close 
to the Poisson value.
The difference of this function to the function $\Psi^{(Ia)}$ is the 
smaller wavelength in regime $III$. 
Here, the boundary roughness is to some extent resolved and the reason for the regular wavefunction lies in the periodicity of the teeth. 
These regular wavefunctions at higher energies should disappear in
systems with random boundary roughness.
Figure~\ref{fig4}(e) (regime $IV$ on the Fig.~\ref{fig4}(f)) represents an interesting intermediate case of a function that is chaotic as well as localized.
$I_0$ in this case is $0.667$, rather close to the Wigner value.
The function $\Psi^{(IV)}$ looks random in the inner rectangular part of 
the billiard, but its amplitudes in this part are very small. 
The largest amplitudes of the function are localized on the border of the teeth. 
So this case is in some sense intermediate between a localized and a chaotic function. 
This seems to be a weak localization mechanism, where the wave is 
reflected at the boundary roughness and interferes constructively inside the teeth.

Accordingly, we found four characteristic types of eigenfunctions, regular ones, localized ones, chaotic ones and intermediate ones between chaotic and localized. 
All eigenfunctions seem to correspond to energy windows, which can be 
characterized by their corresponding $I_0$-values. 
We now introduce several measures for the eigenfunctions that allow us to distinguish between those cases.

(i) As a first characteristics for the eigenfunction analysis we used the distribution of the amplitudes $P(\psi)$. 
By semiclassical arguments it was conjectured that for classically 
chaotic systems most eigenfunctions are a random superposition of plane 
waves, which leads to an amplitude distribution that is a Gaussian 
function \cite{Berry,McDonald,Simmel},
\begin{equation}
P(\psi)={1\over\sqrt{2\pi}}\,e^{-\psi^2/2}
\label{AmDChaos}
\end{equation}
where $\psi=\sqrt{A}\Psi$ is normalized according to 
$\int\left|\psi(x,y)\right|^2dxdy=A$ with the area $A$ of the billiard. This
normalization allows us to compare eigenfunctions of systems with different sizes.
The amplitude distribution of eigenfunctions of a rectangular
billiard, on the other hand, was shown to be \cite{shig94}:
\begin{equation}
P(\psi)= \cases
{{4 \over {\pi^2(2+\psi^2)}}K\left({2-|\psi|\over2+|\psi|} \right), &  $0<|\psi|\le2$; \cr
0,& $|\psi|>2$ ; \cr}
\label{AmDRect}
\end{equation}
where $K(k)$ is the complete elliptic integral of the first kind.
$P(\psi)$ was already studied for the $\pi/3$-rhombus billiard with $g=2$ 
\cite{BiswasJain} and for systems with a pointlike scatterer \cite{shig94}.

In Fig.~\ref{fig5} we show the amplitude distributions of our
eigenfunctions from Fig.~\ref{fig4} by the open circles.
(The filled circles will be explained in the next section.) 
The limiting cases of the Gaussian distribution for random functions and of Eq.~(\ref{AmDRect}) for regular functions are indicated by a dotted and a solid line, respectively.
For the regular looking functions $\Psi^{(Ia)}$ and $\Psi^{(III)}$ 
(cf. Figs.~\ref{fig5}(a) and (d)) we find a very good agreement with Eq.~(\ref{AmDRect}). 
The amplitude distribution on Fig.~\ref{fig5}(b) for function $\Psi^{(Ib)}$ on the other hand consists of one large peak at $\psi \approx 0$. 
This function is localized and only in a very small region of the billiard the amplitude is large. 
The function $\Psi^{(II)}$ on Fig.~\ref{fig5}(c) represents a function with a Gaussian distribution of the amplitude (cf. Eq.~(\ref{AmDChaos})).
This is in line with our estimation that the function looks chaotic. 
The last case of the eigenfunction $\Psi^{(IV)}$ that looks intermediate shows a distribution, which lies between the curves of the Gaussian and the localized function 
(cf. Fig.~\ref{fig5}(e)).

(ii) A second quantity to characterize the eigenfunctions is the 
localization volume $V_{\rm{loc}}^{(\alpha)}$ (participation ratio) 
\cite{Wegner},
\begin{equation}
V_{\rm{loc}}^{(\alpha)}={V_\alpha \over A} \equiv \frac{1}
{A\,\int\left|\Psi^{(\alpha)}\right|^4\,dx dy},
\end{equation}
where $\Psi$ is normalized according to $\int\left|\Psi(x,y)\right|^2dxdy=1$
and $V_\alpha = \left(\int \left|\Psi^{(\alpha)}\right|^4\,dx dy\right)^{-1}$.
For some specific examples of $\Psi(x,y)$, we find
\begin{equation}
V_{\rm{loc}}= \cases
{1 & for constant functions \cr
4/9 & for regular sine or cosine-functions \cr
1/3 & for Gaussian functions \cr
1/A & for $\delta$-functions. 
}
\end{equation}

For the functions of Figs.~\ref{fig4}(a-e), we find 
$V_{\rm{loc}}^{(Ia)}\approx 0.429$, $V_{\rm{loc}}^{(Ib)}\approx 0.025$, 
$V_{\rm{loc}}^{(II)}\approx 0.332$, $V_{\rm{loc}}^{(III)}\approx 0.441$ and 
$V_{\rm{loc}}^{(IV)}\approx 0.060$. 
$V_{\rm{loc}}^{(Ia)}$ as well as $V_{\rm{loc}}^{(III)}$ is very close to the 
value of $4/9$ of a regular cosine-function, whereas $V_{\rm{loc}}^{(Ib)}$ 
is very small and corresponds to a localized state.
The localization volume $V_{\rm{loc}}^{(II)}$ is very close to $1/3$ for
Gaussian functions, which confirms that the function is chaotic.
For the last function, $V_{\rm{loc}}^{(IV)}$ is again small, which means 
that also this function is rather localized, even if its $V_{\rm{loc}}$ is 
slightly larger than $V_{\rm{loc}}^{(Ib)}$. 
In all cases, the values of $V_{\rm{loc}}^{(\alpha)}$ match very well to 
the amplitude distributions $P(\psi)$ of Fig.~\ref{fig5}.

(iii) As a third measure for the eigenfunctions, we investigate the energy 
surface. 
In order to do so, we expand their amplitudes $C_{n,m}^{(\alpha)} =
\lan\Psi^{(\alpha)}\vert\Phi_{n,m}\ran$ in the basis $n,m$ of a rectangular 
billiard, which has the same linear extensions $x,y$ as our rough billiard. 
Here, $\alpha$ enumerates the eigenstates of the rough billiard, while 
$\Phi_{n,m}$ are the eigenfunctions of the rectangular system. 
A similar analysis for chaotic billiards has been performed in 
Refs.~\cite{Frahm2,Yuri}.

The amplitudes $|C_{n,m}^{(\alpha)}|$ of our eigenfunctions are shown in 
Fig.~\ref{fig6}. 
Except for the localized state $\Psi^{(Ib)}$, where the functions 
$\Phi_{n,m}(x,y)$ do not form a good basis, the values of 
$|C_{n,m}^{(\alpha)}|$ appear as peaks that are situated very close to the 
line of constant energy, $E_\alpha\sim (n^2+m^2)$. 
This means that those eigenstates $\Phi_{n,m}(x,y)$ of the rectangular system, 
which have the energy $E_{n,m}$ close to $E_\alpha$,
interfere and form the eigenstates $\Psi^{(\alpha)}$ of the rough system.
However, the number of the participating states is very different. 
The regular states are represented in $n,m$-space by one large peak (one 
coefficient $C_{n,l}^{(\alpha)}$ has the absolut value close to 1), while 
the contribution of the others is vanishing.
This is the case for the functions $\Psi^{(Ia)}$ and $\Psi^{(III)}$, whose 
$n,m$-space can be seen in Figs.~\ref{fig6}(a) and (c), respectively. 
$|C_{n,m}^{(II)}|$ of the chaotic state $\Psi^{(II)}$ is shown in 
Fig.~\ref{fig6}(b).

Here, we find many peaks of roughly equal heights along the energy 
surface, which shows that this state is a superposition of many different 
states $\Phi_{n,m}(x,y)$ that contribute roughly by equal weight and is 
therefore spread over the whole energy surface in $n,m$-space.
A more complicated intermediate situation can be found in 
Fig.~\ref{fig6}(d) (corresponding to the amplitude of $\Psi^{(IV)}$). 
Here, we find a dense distribution of peaks on the energy surface (as 
for the chaotic state), but also several secondary broader peaks that 
are not lying close to the curve of $E_\alpha\sim (n^2+m^2)$ 
and seem to be due to the
high amplitudes localized in the "teeth", which cannot be
described in the basis of the rectangular billiard.
This state is obviously intermediate between localized and chaotic.
Accordingly, also the $n,m$-space matches very well to the preceeding 
measures of the eigenfunctions and to the values of $I_0$.

\section{Distribution of eigenfunctions}

We have seen in the last section that for five selected functions, all measures of the eigenfunctions match very well with the $I_0$-results found from level statistics. 
In intervals, where the level statistics is closer to Poisson-like
behavior ($I_0$ close to $1$), we found eigenfunctions, which are either 
regular or localized. 
In intervals where the level statistics is close to Wigner-like behavior
($I_0\approx 0.637$) on the other hand, the eigenfunctions seem to be chaotic with a Gaussian distribution of the amplitudes and spread over the whole energy surface in $n$,$m$-space. 
Now, we look at the distribution of the eigenfunctions inside a given energy interval.
For this purpose, we calculated more than $100$ eigenstates for each of the four energy regimes of Fig.~\ref{fig4}(f) ($341$ eigenfunctions in regime $I$, $131$ states in regime $II$, $139$ states in regime $III$ and $144$ in regime $IV$). 
In each case, we calculated $V_{\rm{loc}}$ and the amplitude distribution $P(\psi)$ over all eigenfunctions. 

In Fig.~\ref{fig7}, we show the normalized histograms of the 
$V_{\rm{loc}}$-values for the four energy regimes.
In the first regime (see Fig.~\ref{fig7}~(a)), we find two peaks:
one peak is close to $V_{\rm{loc}}=0.44$, which corresponds to eigenstates of a rectangular billiard, and one peak at $V_{\rm{loc}}<0.1$, indicating localized states. 
In the second energy regime (see Fig.~\ref{fig7}~(b)), we find a narrow distribution of $V_{\rm{loc}}$ around the value of $0.33$ of chaotic functions, in good agreement with the relatively low value of $I_0=0.688$ in this regime.
In the third regime (see Fig.~\ref{fig7}~(c)), the values of $V_{\rm loc}$ are distributed in the interval $[0.35-0.46]$ with two peaks around $0.37$ and $0.44$. 
That indicates basically regular states with a slight end towards
random behavior, which is in line with the value of $I_0=0.893$ in this regime.
In the fourth energy regime (see Fig.~\ref{fig7}~(d)), most of the 
$V_{\rm{loc}}$-values are roughly distributed in the interval $[0.1,0.2]$, which means again that the states are quite localized.
Nevertheless the value of $I_0=0.667$ in this regime is rather close to $I_0^{\rm Wigner}$ for chaotic functions.
This contradiction accounts for intermediate states between localized and chaotic.

Next, we look at the average amplitude distributions in our four regimes, 
which are shown in Fig.~\ref{fig5}~(a-e) by the full circles. 
Since in the regime $I$ two types of eigenfunctions exist, we split them 
into two groups according to their $V_{\rm{loc}}$-values. 
The average amplitude distribution for 263 states with $V_{\rm{loc}}>0.2$ 
is presented in Fig.~\ref{fig5}(a) and the one for 78 functions with 
$V_{\rm{loc}}\le 0.2$ on Fig.~\ref{fig5}(b). 
The distributions in the regimes $II-IV$ are shown in 
Fig.~\ref{fig5}~(c-e) by full circles.
The distributions are very close to those for the individual functions 
(open circles), from the preceeding section, showing that the features, which 
we found for 
the single functions are characteristic for a whole energy regime. 
The eigenfunctions of regime $I$ are either localized or regular, the eigenfunctions of regime $II$ chaotic, the eigenfunctions of regime $III$ again close to regular functions and in regime $IV$, there are a lot of 
intermediate functions between localized and chaotic.
It is remarkable that even the intermediate functions cover a whole energy window and explains the discrepancy between the rather small values of $V_{\rm{loc}}$ and the value of $I_0$, which is close to the Wigner-limit.

With these measures in mind, we can finally look at the systems with a random distribution of teeth widths, where $I_0$ approached the Wigner limit very fast.
In Fig.~\ref{fig8}, we show (a) the histogram of the $V_{\rm{loc}}$-values and in (b) the amplitude distribution of $152$ eigenfunctions of system $R_1$ in the energy interval $[0.225-0.240]$.
The values of the localization volumes show a narrow distribution around the value of $V_{\rm{loc}}=0.33$ of Gaussian functions and the amplitude
distribution $P(\psi)$ coincides with a Gaussian function.
So, these functions are quite obviously a random superposition of plane waves.

\section{Conclusions}

In summary, we investigated the level statistics and the eigenfunctions of 
pseudointegrable rough billiards with high genus numbers $g$. 
They had a rectangular body and a rough boundary with small 
''teeth'' of different widths, distances and heights. 
The level statistics was found to be intermediate between 
Poisson- and Wigner-like behavior and approached the latter with increasing
$g$. Especially for small level distances, the $\Delta_3$-curves can come
very close to the Wigner statistics.
This behavior was similar to the one of Seba billiards
with an increasing number of pointlike scatterers \cite{shig94,chsh96}.

Additionally, we found different energy intervals 
with characteristic types of eigenfunctions, which correspond to a 
special behavior of the level statistics. 
In order to classify the eigenfunctions we employed several measures: 
the localization volume $V_{\rm{loc}}$, the amplitude distribution $P(\psi)$
and the behavior of the eigenfunctions in $n,m$-space.

We found that all systems have a low-energy regime, where the wavelengths of 
the eigenfunctions are too large to resolve the boundary teeth and 
thus only see the main rectangular body of the billiard. 
The eigenfunctions in the low-energy regime have the characteristics of 
regular functions with the level statistics close to the Poisson 
distribution.
However, for systems with a random distribution of the boundary teeth or 
for very broad teeth, this regime stays very small and the high-energy 
regime is quickly reached.

For higher energies, the eigenfunctions of
systems with either random boundary roughness or with not too small teeth, 
are characterized by a close-to-Gaussian amplitude 
distribution $P(\psi)$ with a localization volume of $V_{\rm{loc}}\approx 0.33$.
They can be constructed by a random superposition of many regular functions.  
The second half moments $I_0$ of the energy spacing distribution $p(s)$
at higher energies are -- apart from fluctuations -- energy-independent and 
close to a Wigner-like behavior for large $g>20$. 
Also the spectral rigidity $\Delta_3$ changes from Poisson-like towards 
Wigner-like behavior when $g$ increases. However, $\Delta_3$ changes slower 
and still shows deviations from the Wigner behavior even for large $g$.

For the case of periodic and narrow boundary teeth, we find many energy 
regimes, where the level statistics is close to Poisson behavior.
We find that those functions are either localized with very small values 
of $V_{\rm{loc}}$ or regular cosine functions with $V_{\rm{loc}}$ close to 
$0.44$. 
Also an interesting regime occurs, where the eigenfunctions show intermediate 
behavior with chaotic traces in the main body of the system but very 
high amplitudes inside the teeth, indicating weak localization.

Accordingly, deviations from a Wigner-like behavior of the level 
statistics are due to three different effects, (i) a poor resolution of 
the boundary roughness at very small energies, (ii) regular wavefunctions
 due to periodic boundary roughness and (iii) localization. 
However, for random boundary roughness and not too small energies, those 
effects are not pronounced and the eigenfunctions can mostly be described 
by a random superposition of plane waves, which is for pseudointegrable
systems remarkable.

For further research it would be interesting to investigate the 
eigenfunctions in the high-energy regime, which might help to understand 
the $\Delta_3$ data for large values of $g$. 
As $\Delta_3$ measures the long-range correlations between eigenvalues it 
seems likely that even a very small amount of regular or localized
eigenfunctions in an energy interval of otherwise chaotic functions can 
lead to these deviations.

\section{Acknowledgements}
We gratefully acknowledge financial support from the Deutsche 
Forschungsgemeinschaft. We would like to thank Jan Kantelhardt for
making us familiar with the Lanczos algorithm and the calculation of
the second half moments and Armin Bunde for a careful reading of the
manuscript and interesting remarks.

\newpage
\begin{table}  
\begin{tabular}{|c|p{1cm}|p{1cm}|p{1cm}|p{1cm}|p{1cm}|p{1cm}|p{1cm}|}
\hline
&&&&&&&\\
Geometry & $a$ & $h$ & $b_x$ & $b_y$ & $N_x$ & $N_y$ & $g$\\ 
&&&&&&&\\\hline
 $A_1$ & 8 & 8 & -   & - & 1 &  0 & 3\\
 $A_2$ & 8 & 8 & -  & 150 & 1 &  2 & 5\\
 $A_3$ & 8 & 8 & 117 & 150 & 3 &  2 & 9\\
 $A_4$ & 8 & 8 & 55 &   71 & 6 &  4 & 19\\
 $A_5$ & 8 & 8 & 39 &   37 & 8 &  7 & 29\\
 $A_6$ & 8 & 8 & 26 &   27 & 11 & 9 & 39\\
 $A_7$ & 8 & 8 & 21 &   18 & 13 & 12 & 49\\
\hline
\end{tabular}
\caption[]{\small Table of the geometries
used in Fig.~2. The parameters $a$, $h$, $b_x$ and $b_y$ refer
to the ones given in Fig.~\ref{fig1}. The values of the parameters 
are given in units of the lattice constant $d$.
}\label{table1}
\end{table}
\begin{table}  
\begin{tabular}{|c|p{2cm}|p{2cm}|p{2cm}|p{2cm}|}
\hline
&&&&\\
Geometry & $a$, $b_x$, $b_y$, $h$ & $N_x$ & $N_y$ & $g$\\ 
&&&&\\\hline
 $R_1$ & $[4,15]$ & 22 &  17 & 77\\
 $R_2$ & $[4,10]$ & 29 &  22 & 101\\
\hline
\end{tabular}
\caption[]{\small Table of the series of the geometries used in Fig.~\ref{fig3}. 
The parameters $a$, $b_x$, $b_y$, $h$ have random values from the indicated intervals. 
The values of the parameters are given in units of the lattice constant $d$.}
\label{table2}
\end{table}
\begin{table}  
\begin{tabular}{|c|p{2cm}|p{2cm}|p{2cm}|p{2cm}|p{2cm}|}
\hline
&&&&&\\
Geometry & $a=b_x=b_y$ & $h$ & $N_x$ & $N_y$ & $g$\\ 
&&&&&\\\hline
 $B_1$ &  4 &  4 &  48 &  40 & 87\\
 $B_2$ &  4 &  8 &  48 &  40 & 87\\
 $B_3$ &  4 & 16 &  48 &  40 & 87\\
\hline
 $B_4$ &  8 &  4 &  24 &  20 & 43\\
 $B_5$ &  8 &  8 &  24 &  20 & 43\\
 $B_6$ &  8 & 16 &  24 &  20 & 43\\
\hline
\end{tabular}
\caption[]{\small Table of the series of the geometries
used in Fig.~\ref{fig3}. The parameters $a$, $h$, $b_x$ and $b_y$ refer
to the ones given in Fig.~\ref{fig1}. In both series of the geometries 
the width and distance between the teeth remain the same while the height is 
changing. The values of the parameters are given in units of the lattice 
constant $d$.}
\label{table3}
\end{table}

\unitlength 3.7mm
\vspace*{0mm}
\begin{figure}
\begin{picture}(0,30)
\def\epsfsize#1#2{1.4#1}
\put(0,0){\epsfbox{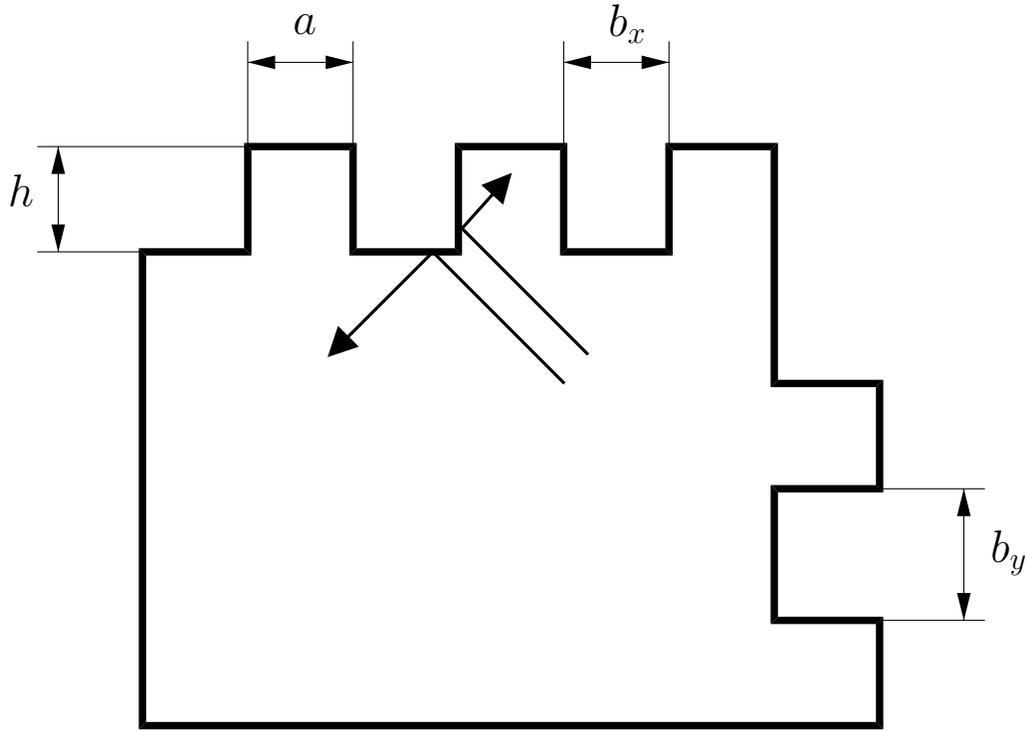}}
\put(2,19){\makebox(1,1){{\LARGE $h$}}}
\put(12.2,25){\makebox(1,1){{\LARGE $a$}}}
\put(23.8,25){\makebox(1,1){{\LARGE $b_x$}}}
\put(37.5,6){\makebox(1,1){{\LARGE $b_y$}}}
\end{picture}
\caption[]{\small Sketch of the considered pseudointegrable geometry. 
The parameter $a$ is a width of the "teeth", $h$ is their height,
$b_x$ and $b_y$ are the distances between them in $x$- and $y$-directions. 
$N_x$ and $N_y$ are the numbers of "teeth" in $x$- and $y$-directions. The 
genus number $g$ of this geometry is $g=1+G_i$, where $G_i$ is the number of
salient corners with angles of $3\pi/2$. 
The beam splitting property of these corners is shown by the two arrows 
that indicate two different trajectories of classical particles.
}
\label{fig1}
\end{figure}

\newpage
\unitlength 4.0mm
\vspace*{0mm}
\begin{figure}
\begin{picture}(0,30)
\def\epsfsize#1#2{0.7#1}
\put(5,1){\epsfbox{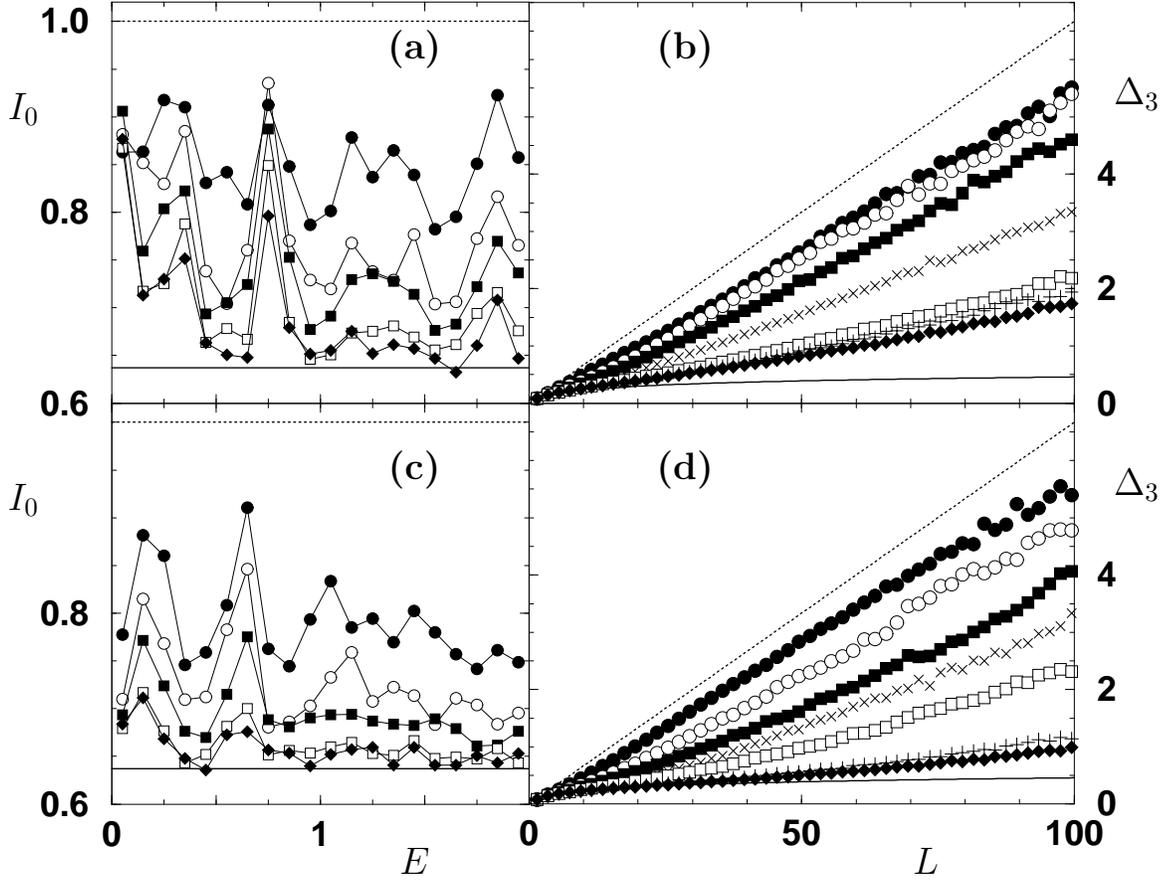}}
\put(0,25){\makebox(2,2){{\Large \bf $I_0$}}}  
\put(0,12){\makebox(2,2){{\Large \bf $I_0$}}} 
\put(13,0){\makebox(2,2){{\Large \bf $E$}}} 
\put(37,12.5){\makebox(2,2){{\Large \bf $\Delta_3$}}} 
\put(37,25.5){\makebox(2,2){{\Large \bf $\Delta_3$}}} 
\put(30,0){\makebox(2,2){{\Large \bf $L$}}} 
\put(13,27){\makebox(2,2){{\Large \bf(a)}}}  
\put(22,27){\makebox(2,2){{\Large \bf(b)}}} 
\put(13,13){\makebox(2,2){{\Large \bf(c)}}}  
\put(22,13){\makebox(2,2){{\Large \bf(d)}}}  
\end{picture}
\caption[]{\small (a) and (c): The second half moments $I_0$ 
are plotted versus the dimensionless energy $E$ for several systems. 
(b) and (d): $\Delta_3(L)$ from the fixed energy interval $E\in[1.5,2.0]$ is 
plotted versus $L$ for the same systems as in (a) and (c). 
Figs.~(a) and (b) refer to Dirichlet, (c) and (d) to Neumann boundary conditions.
The symbols refer to the systems of Tab.~\ref{table1} with 
increasing genus number $g=3$ (full cicles), $g=5$ (open circles), 
$g=9$ (full squares), $g=19$ ($x$), $g=29$ (open squares), $g=39$ ($+$) and 
$g=49$ (full diamonds). In (a) and (c) the systems of $g=19$ and $39$ are
omitted for a better visibility.
$I_0$, $\Delta_3$ and $L$ are dimensionless. In all figures,
the Poisson-limit is indicated by a dotted and the Wigner-limit by a solid line. 
}
\label{fig2}\end{figure}

\newpage
\unitlength 3.5mm
\vspace*{0mm}
\begin{figure}
\begin{picture}(0,40)
\def\epsfsize#1#2{0.9#1}
\put(2,2){\epsfbox{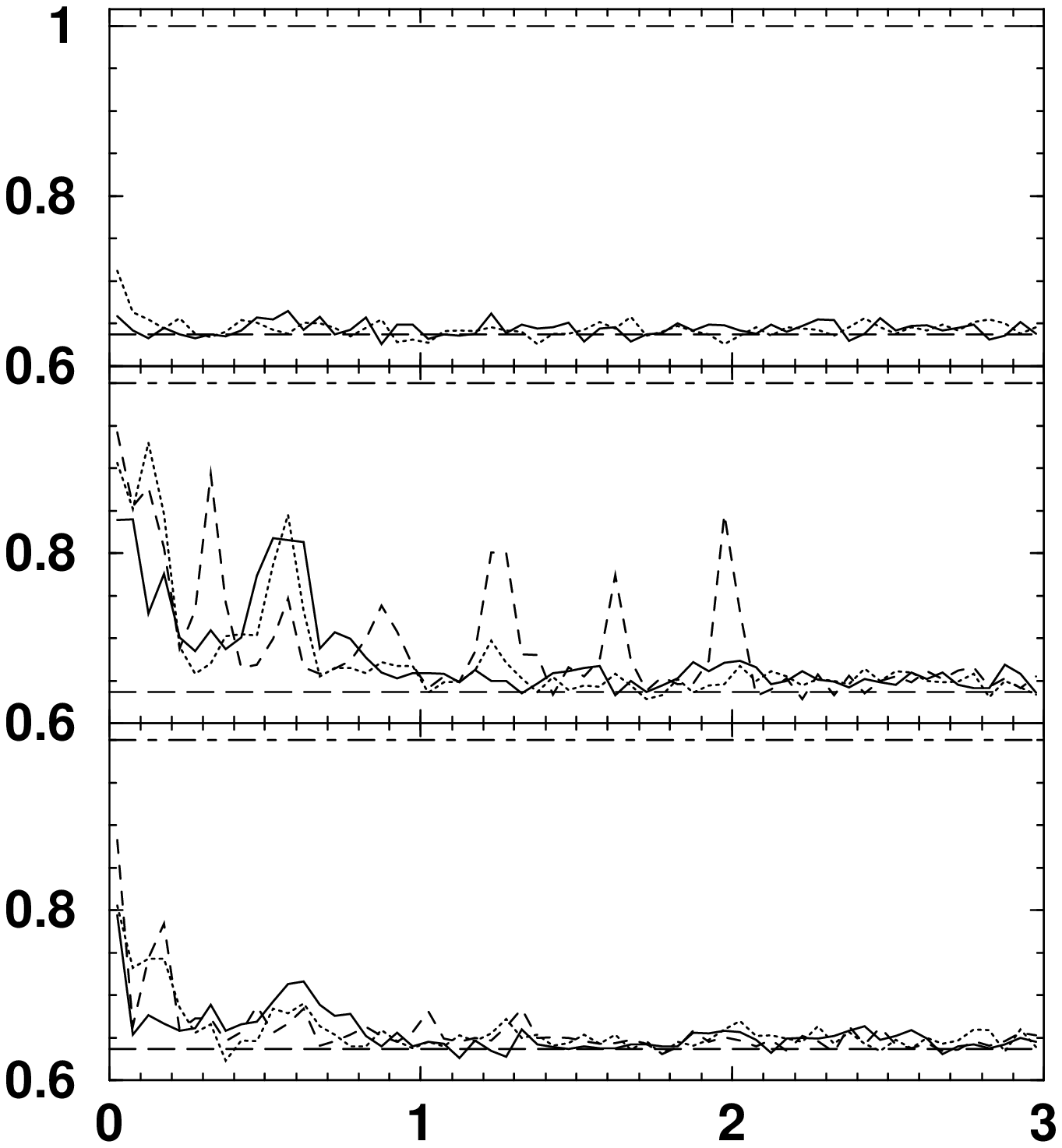}}
\put(33,36){\makebox(2,2){{\Large \bf(a)}}}  
\put(33,24){\makebox(2,2){{\Large \bf(b)}}} 
\put(33,12){\makebox(2,2){{\Large \bf(c)}}}  
\put(0,37){\makebox(2,2){{\Large $I_0$}}}  
\put(0,25){\makebox(2,2){{\Large $I_0$}}}  
\put(0,13){\makebox(2,2){{\Large $I_0$}}}  
\put(20,1){\makebox(2,2){{\Large $E$}}}  
\end{picture}
\caption[]{\small The second half moment $I_0$ is plotted versus the 
dimensionless energy $E$ for Neumann boundary conditions and for different 
geometries (cf. Fig.~\ref{fig1}). 
The systems of (a) have random parameters $a$, $b_x$, $b_y$ and $h$ with
different widths of the distributions (cf. Tab.~\ref{table2}).
The solid line corresponds to the geometry $R_1$, the dotted line for 
geometry $R_2$. 
Figure (b) shows the systems $B_1$, $B_2$ and $B_3$ (cf. Tab.~\ref{table3}), 
which have constant values of $a=b_x=b_y=4$ and an increasing height: $h=4$ 
(solid line), $h=8$ (dotted line), $h=16$ (dashed line). 
Figure (c) shows the systems $B_4$, $B_5$ and $B_6$ (cf. Tab.~\ref{table3}) 
with parameters $a=b_x=b_y=8$ and the height $h=4$ (solid line), $h=8$ 
(dotted line), $h=16$ (dashed line).
}
\label{fig3}
\end{figure}

\newpage
\begin{center}
\unitlength 1.46mm
\vspace*{0mm}
\hspace*{5mm}{
\begin{figure}
\begin{picture}(100,110)
\def\epsfsize#1#2{0.40#1}
\put(0,80){\epsfbox{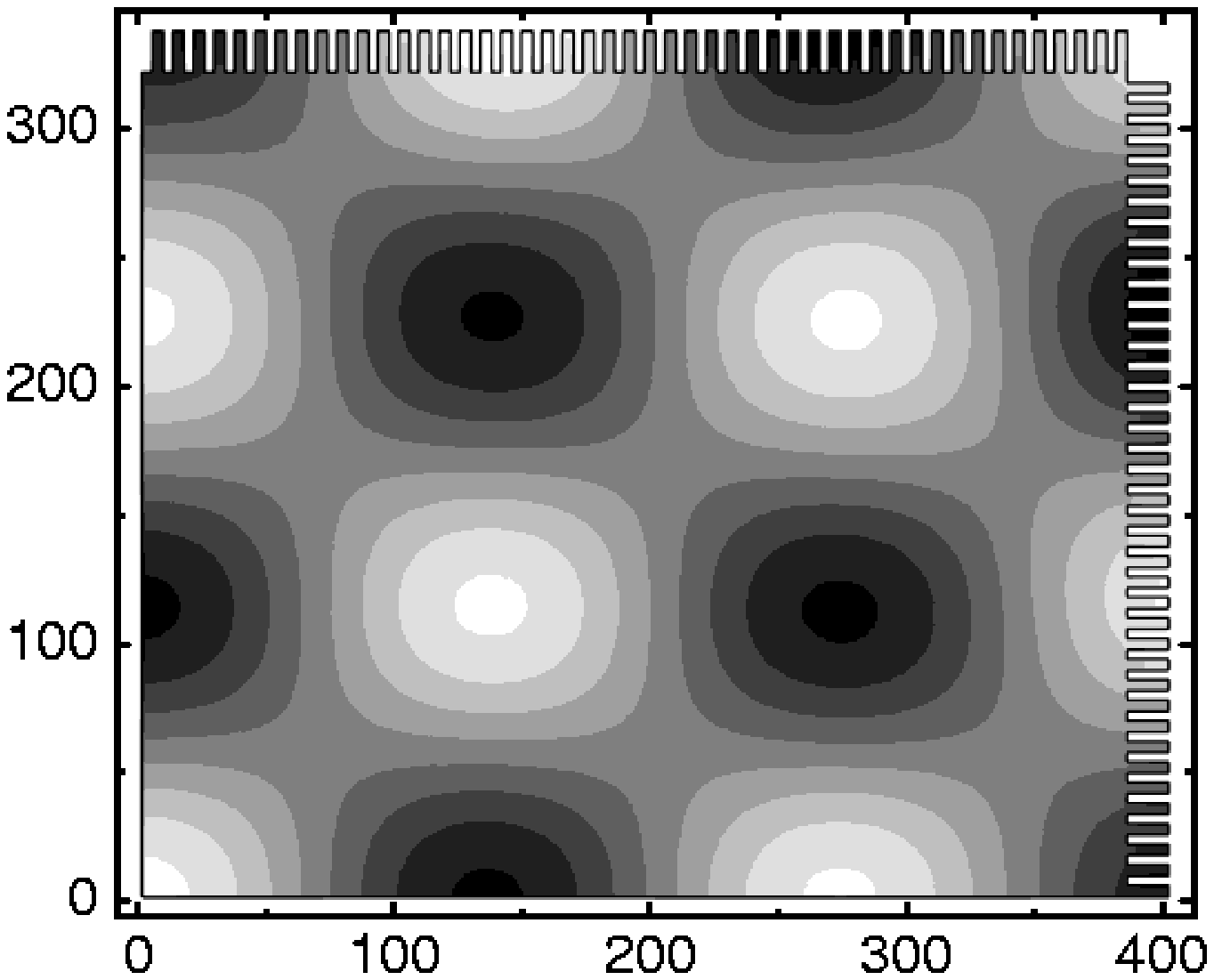}}
\put(50,80){\epsfbox{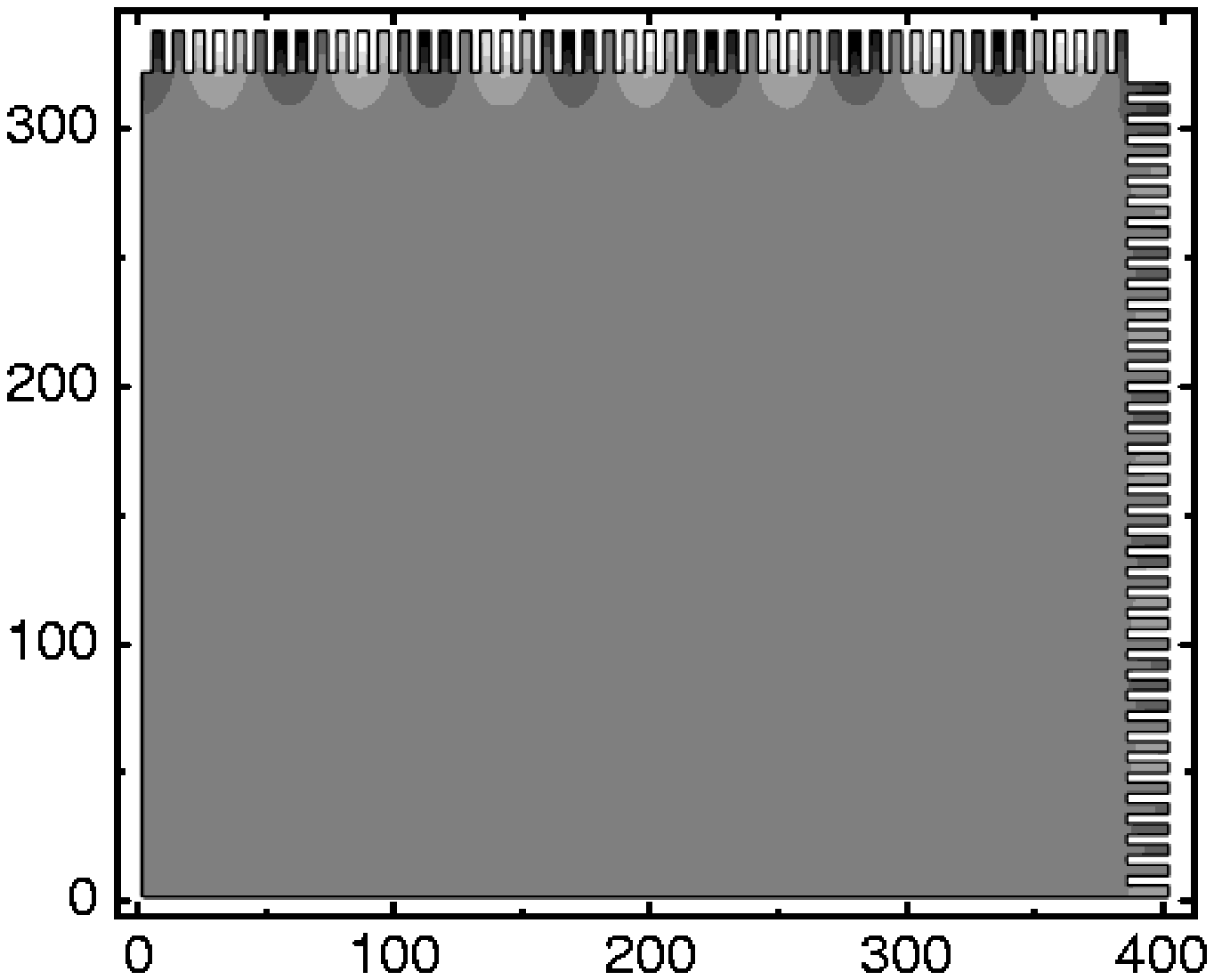}}
\put(0,40){\epsfbox{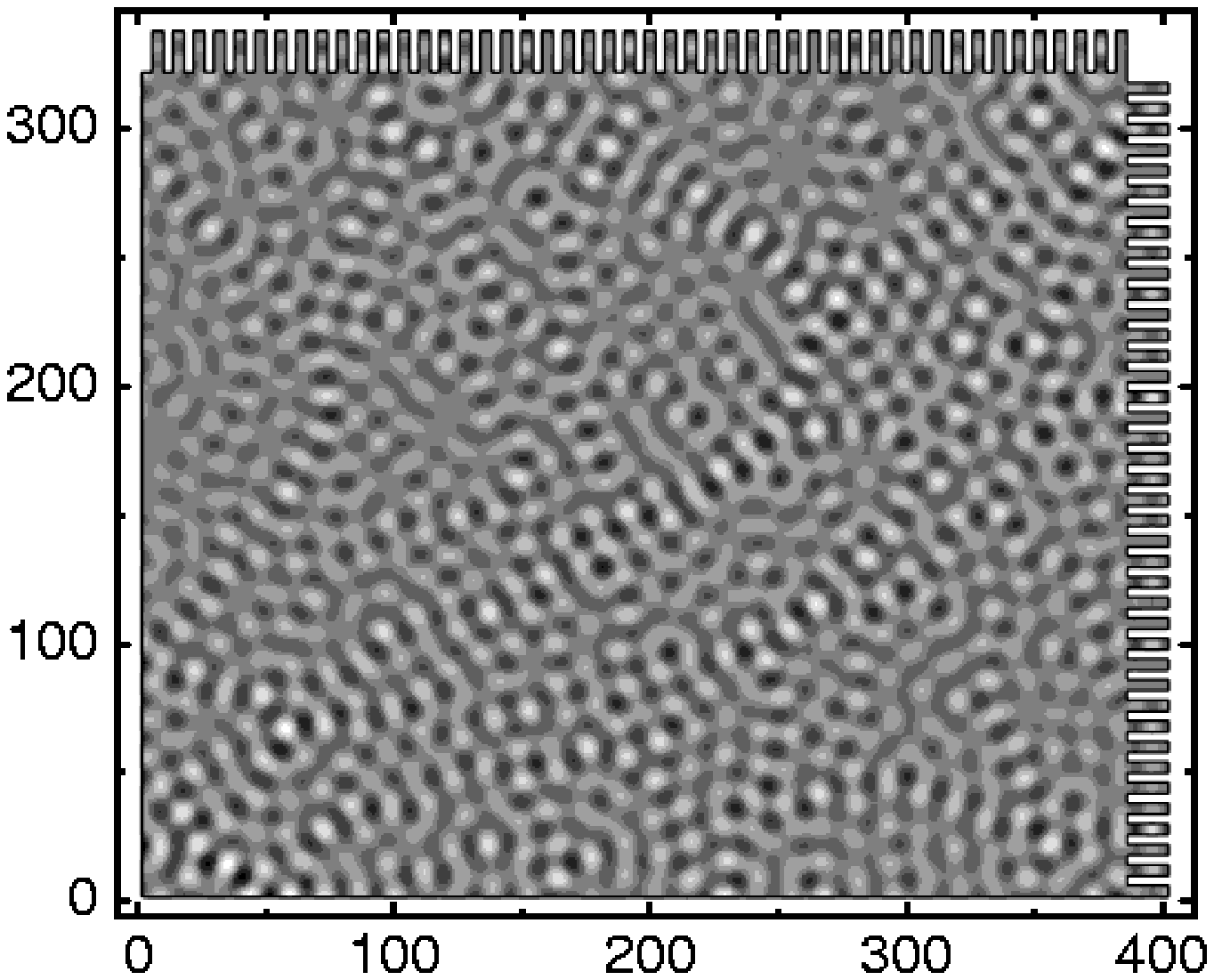}}
\put(50,40){\epsfbox{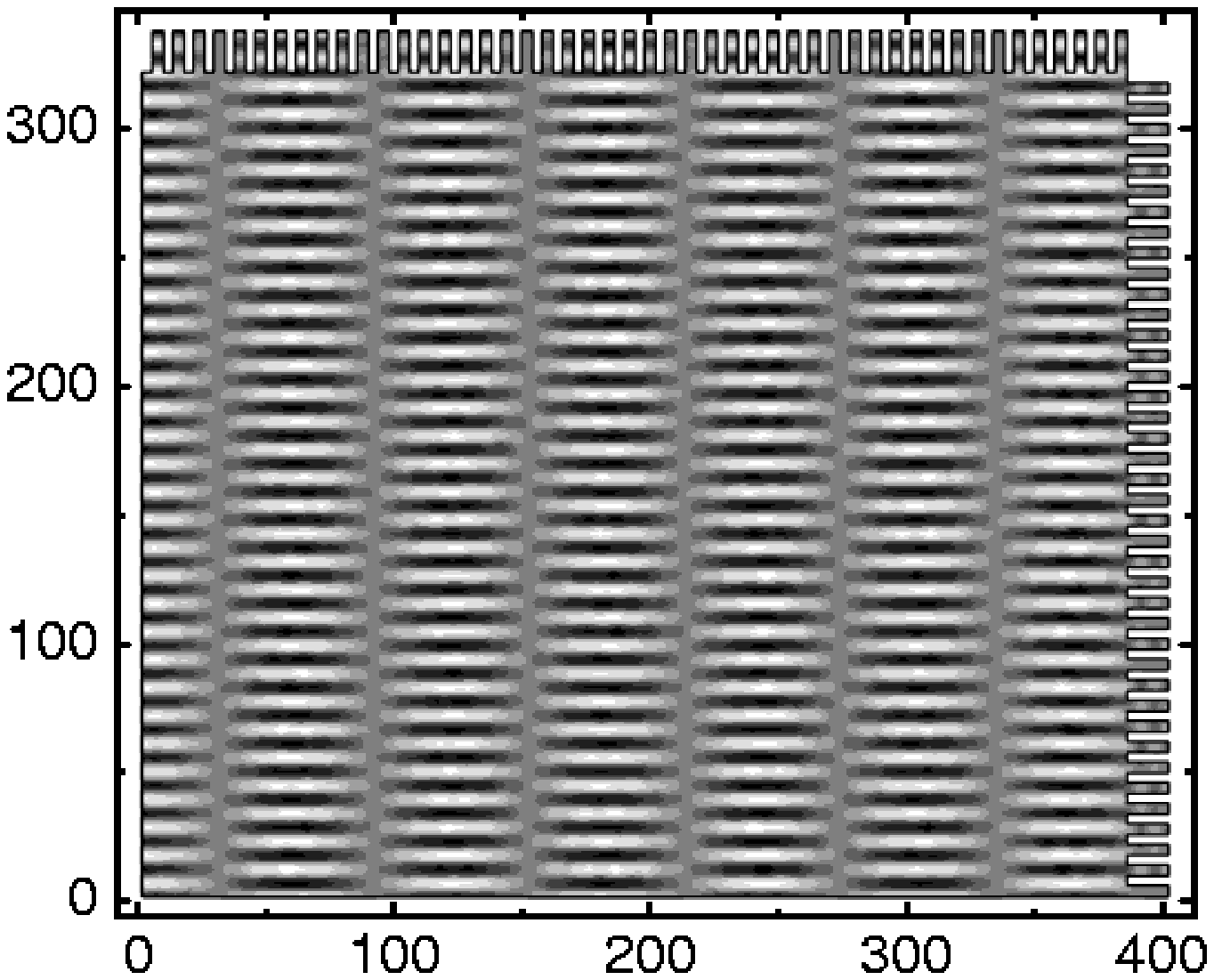}}
\put(0,0){\epsfbox{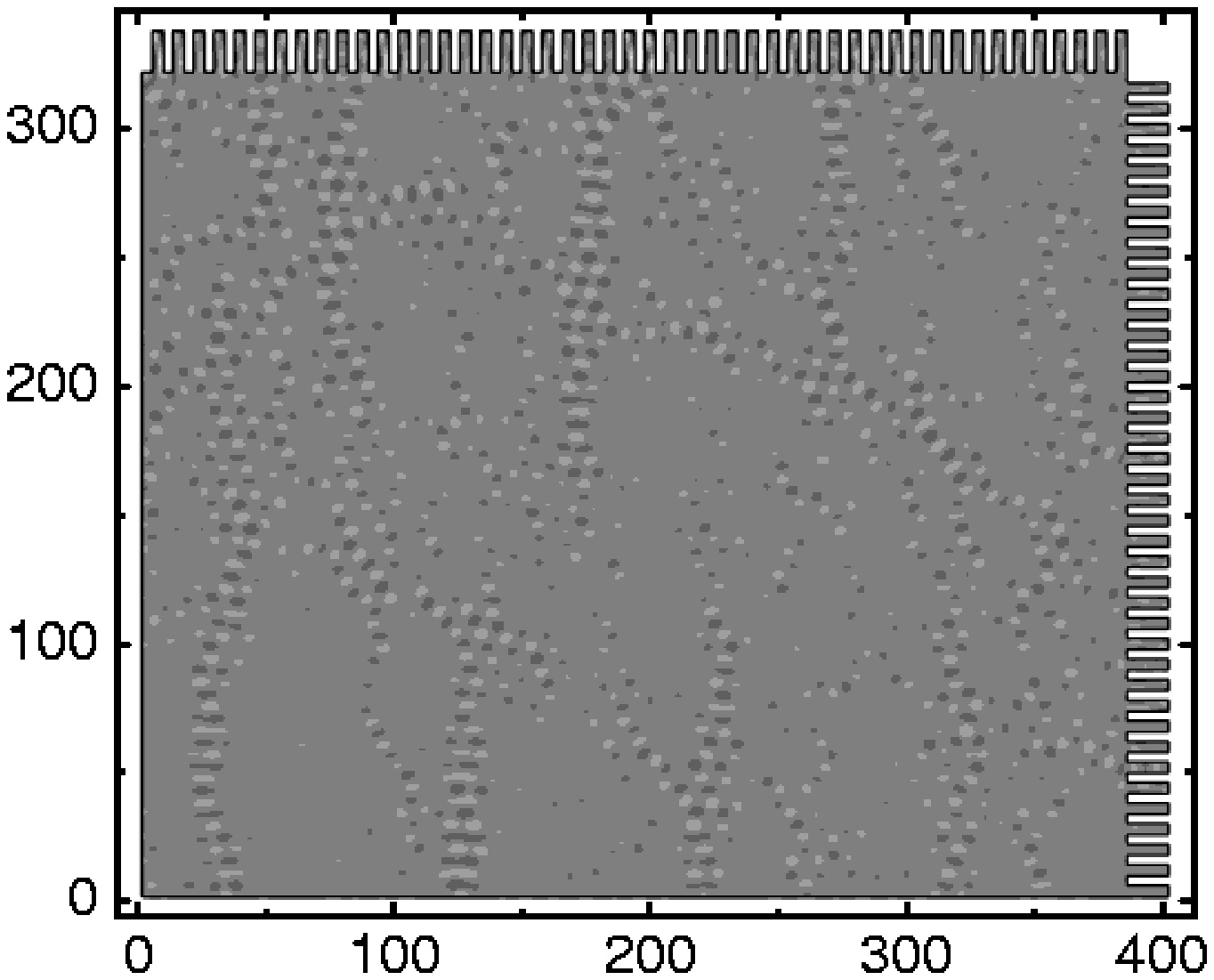}}
\def\epsfsize#1#2{0.348#1}
\put(50,3){\epsfbox{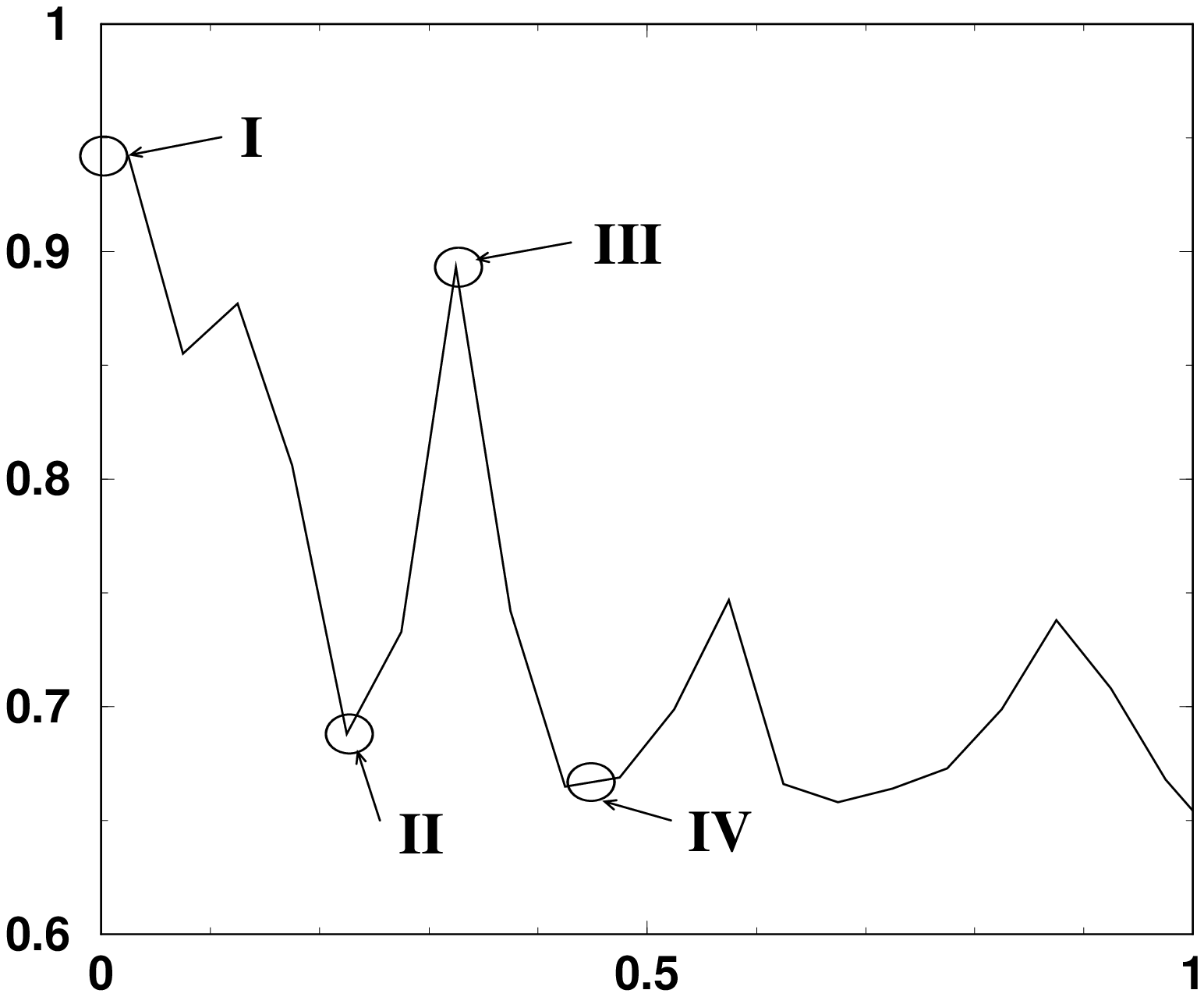}}
\put(49,31){\makebox(1,1){$I_0$}}
\put(79,2){\makebox(1,1){$E$}}
\put(21,116){\makebox(1,1){{\Large $\Psi^{(Ia)}$}}}
\put(71,116){\makebox(1,1){{\Large $\Psi^{(Ib)}$}}}
\put(21,76){\makebox(1,1){{\Large $\Psi^{(II)}$}}}
\put(71,76){\makebox(1,1){{\Large $\Psi^{(III)}$}}}
\put(21,36){\makebox(1,1){{\Large $\Psi^{(IV)}$}}}
\put(40,110){\makebox(1,1){{$\displaystyle (\mathrm{a})$}}}
\put(90,110){\makebox(1,1){{$\displaystyle (\mathrm{b})$}}}
\put(40,70){\makebox(1,1){{$\displaystyle (\mathrm{c})$}}}
\put(90,70){\makebox(1,1){{$\displaystyle (\mathrm{d})$}}}
\put(40,30){\makebox(1,1){{$\displaystyle (\mathrm{e})$}}}
\put(90,30){\makebox(1,1){{$\displaystyle (\mathrm{f})$}}}
\end{picture}
\caption[]{\small (a-e) Typical eigenfunctions from different energy 
windows for the geometry $B_3$. 
The amplitudes are indicated by different gray levels.
The white regions stand for positive amplitudes, the black ones for 
negative amplitudes. 
The neutral gray tone stands for nearly zero amplitude. 
The black contour line shows the border and does not correspond to any 
amplitudes.
In (e), the largest amplitudes lie at the border and are hidden by the 
contour line.
In (f), $I_0$ is plotted versus the dimensionless energy $E$ and the regions, 
from where the functions (a)-(e) are taken, are indicated by circles.
The functions (a) and (b) ($\Psi^{(Ia)}$ and $\Psi^{(Ib)}$) are both taken 
from the region $I$.}
\label{fig4}
\end{figure}}
\end{center}

\newpage
\unitlength 3.5mm
\vspace*{0mm}{
\begin{figure}
\begin{picture}(0,45)
\def\epsfsize#1#2{0.9#1}
\put(0.6,2){\epsfbox{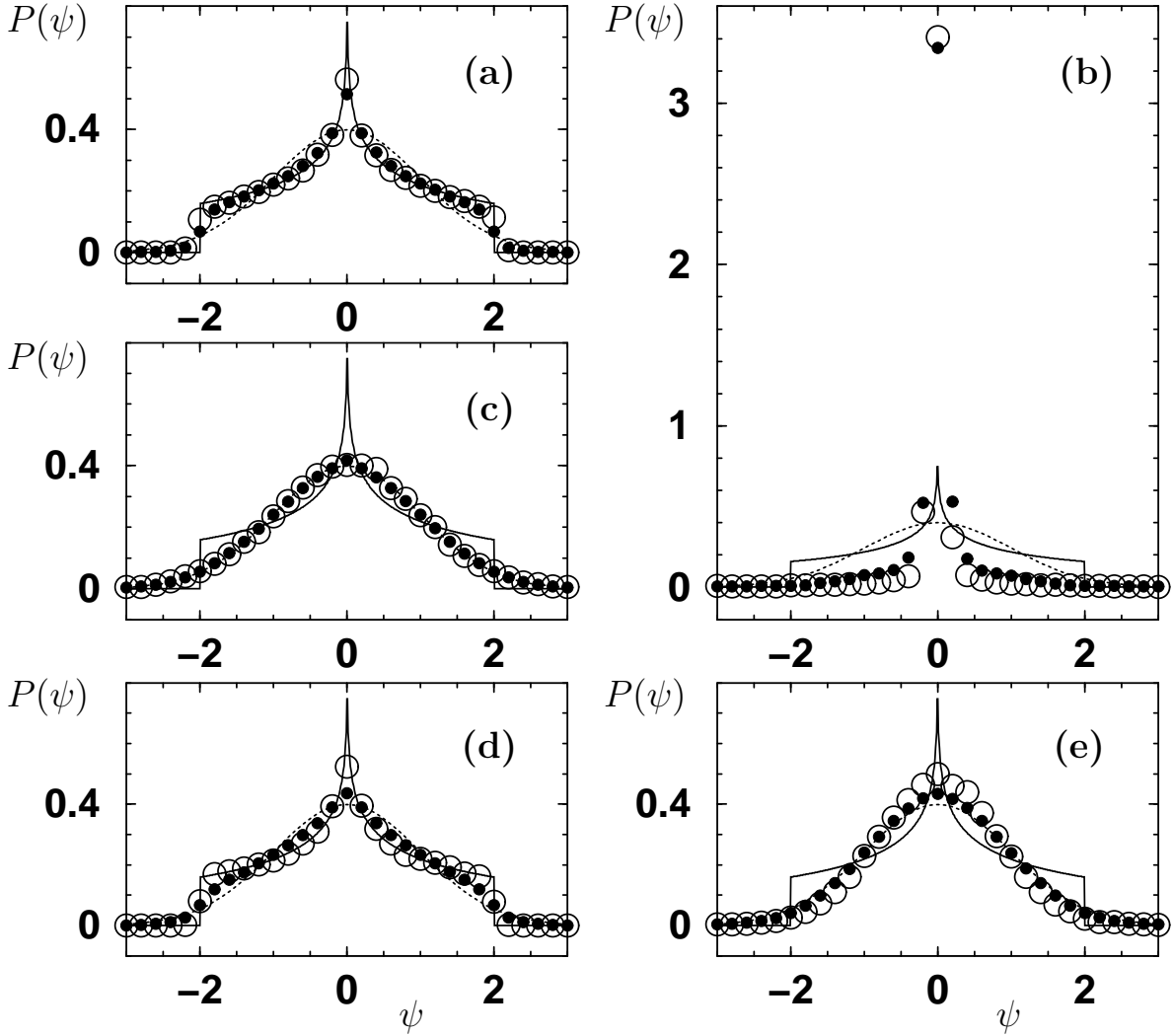}}
\put(0,39){\makebox(2,2){{\Large \bf $P(\psi)$}}}
\put(0,26){\makebox(2,2){{\Large \bf $P(\psi)$}}} 
\put(0,13){\makebox(2,2){{\Large \bf $P(\psi)$}}} 
\put(23,39){\makebox(2,2){{\Large \bf $P(\psi)$}}}
\put(23,13){\makebox(2,2){{\Large \bf $P(\psi)$}}} 
\put(14,0.7){\makebox(2,2){{\Large \bf $\psi$}}} 
\put(37,0.7){\makebox(2,2){{\Large \bf $\psi$}}}
\put(17,37){\makebox(2,2){{\Large \bf (a)}}}
\put(40,37){\makebox(2,2){{\Large \bf (b)}}}  
\put(17,24){\makebox(2,2){{\Large \bf (c)}}}  
\put(17,11){\makebox(2,2){{\Large \bf (d)}}}  
\put(39.7,11){\makebox(2,2){{\Large \bf (e)}}}  
\end{picture}
\caption[]{\small The dimensionless amplitude distribution of (i) the single 
eigenfunctions presented on Fig.~\ref{fig4} (open circles) and 
(ii) the averaged amplitude distribution over many eigenfunctions in the 
corresponding energy intervals as explained in section V (full circles). 
For single eigenfunctions the letters (a)-(e) correspond to $\Psi^{(Ia)}$, 
$\Psi^{(Ib)}$, $\Psi^{(II)}$, $\Psi^{(III)}$, $\Psi^{(IV)}$ respectively.
The dotted line indicates the Gaussian distribution and the solid one the 
distribution of regular $\sin$ or $\cos$-functions (Eq.~\ref{AmDRect}).
Functions (a) and (d) are regular; (b) - localized; (c) - Gaussian 
(chaotic) and (e) - intermediate between chaotic and localized. }
\label{fig5}
\end{figure}}

\newpage
\unitlength 2.1mm
\vspace*{0mm}
\hspace*{5mm}
\begin{center}
\begin{figure}
\begin{picture}(40,80)
\def\epsfsize#1#2{0.32#1}
\put(6,60){\epsfbox{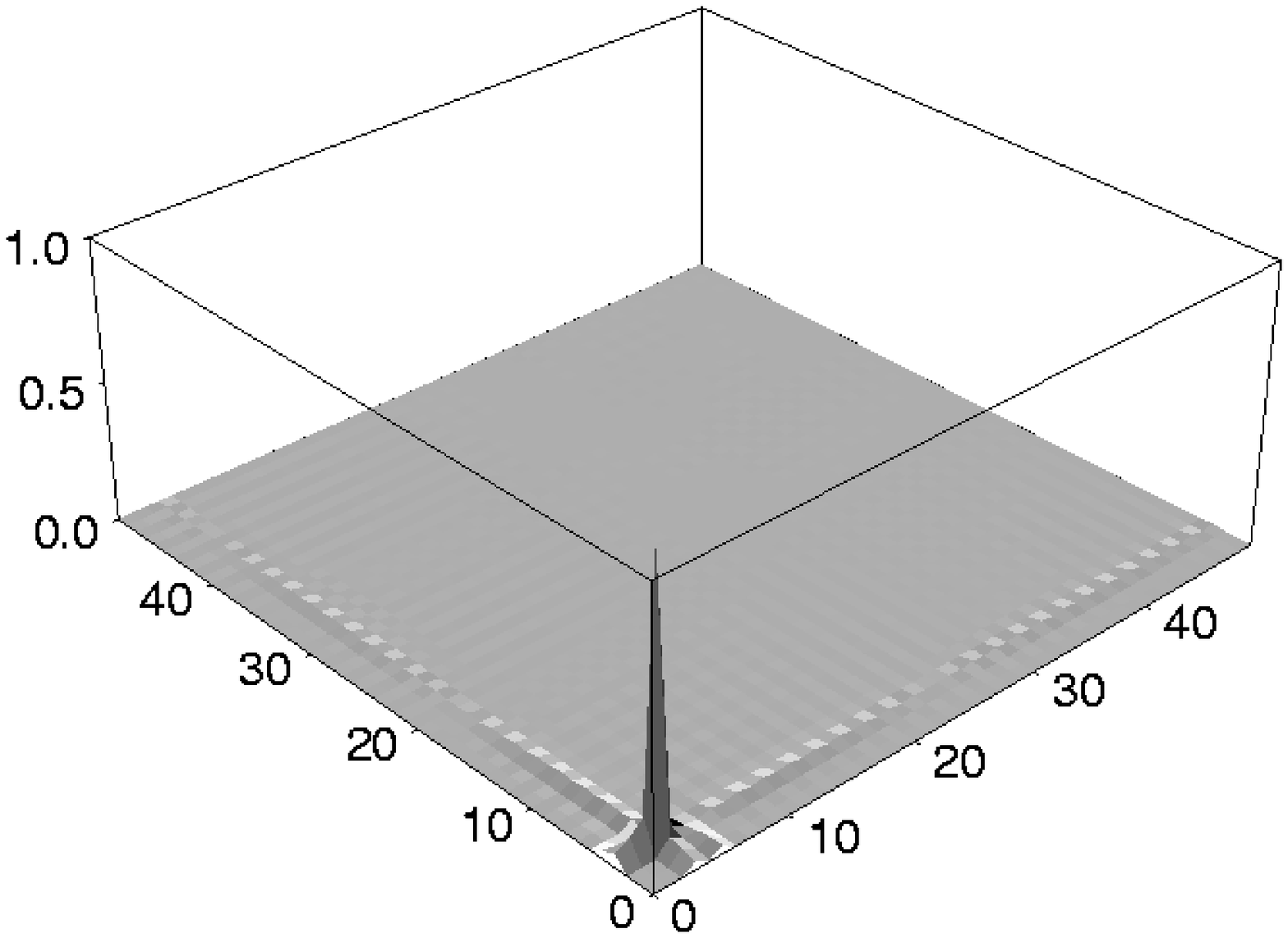}}
\put(6,40){\epsfbox{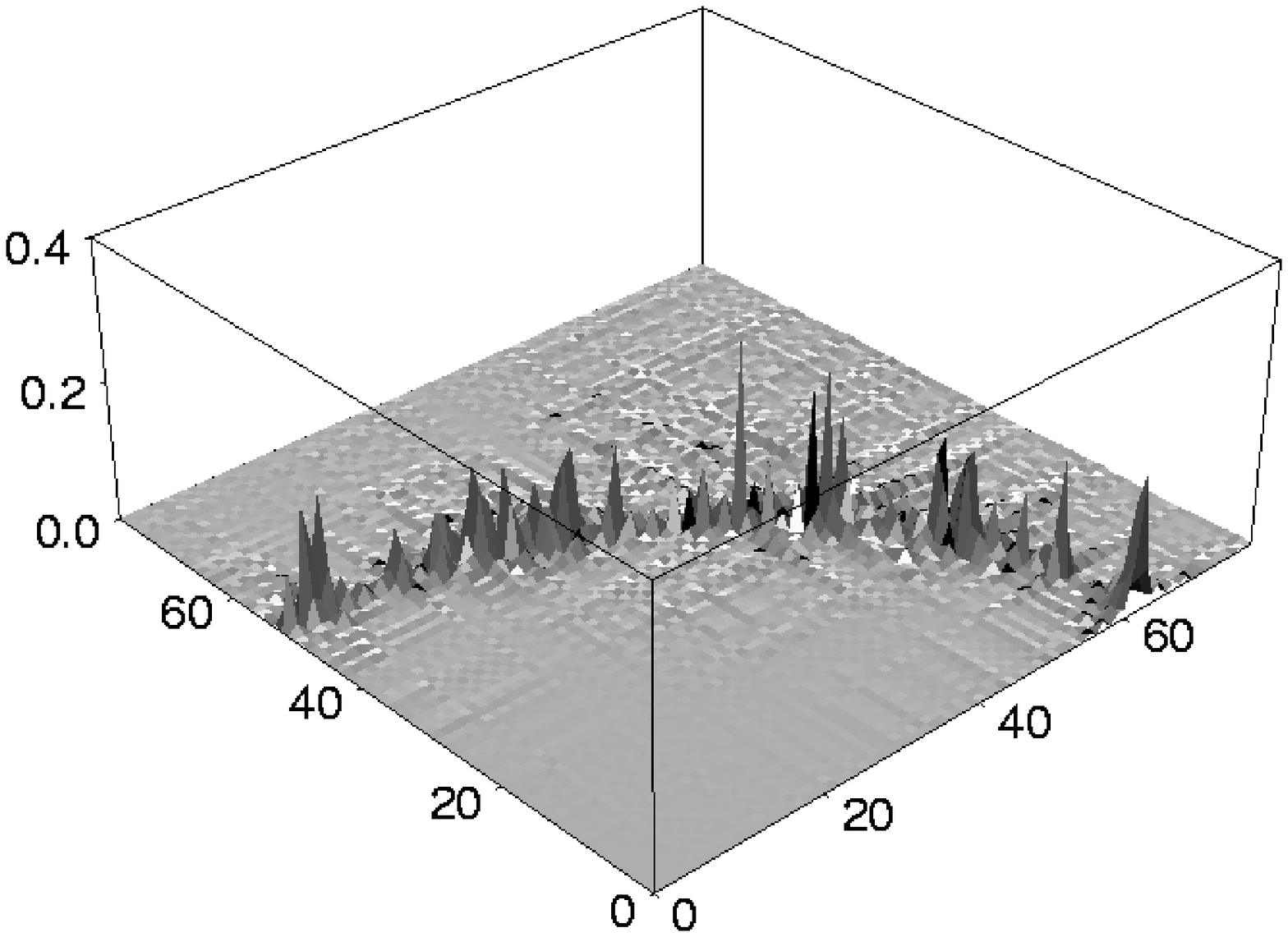}}
\put(6,20){\epsfbox{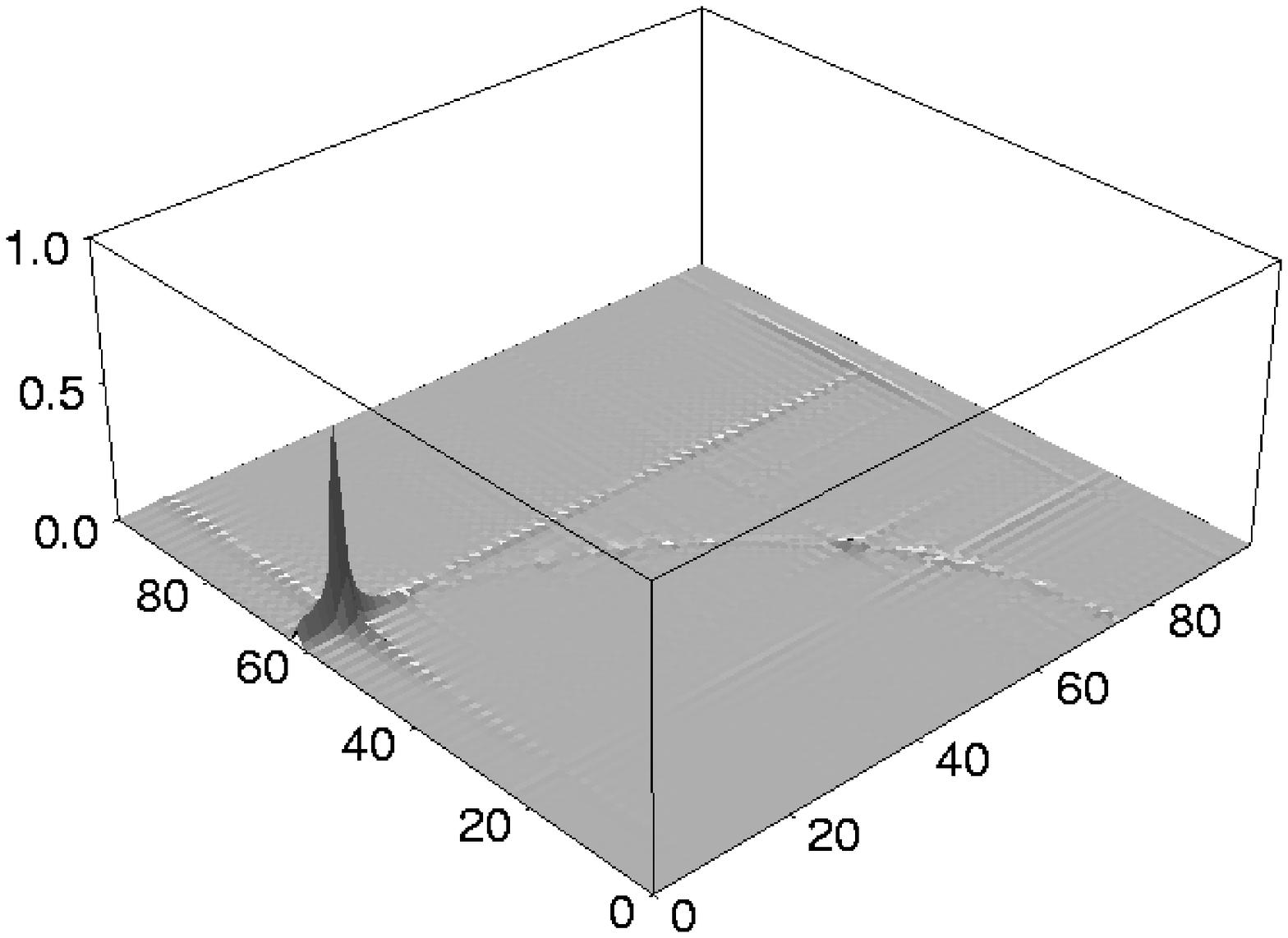}}
\put(6,0){\epsfbox{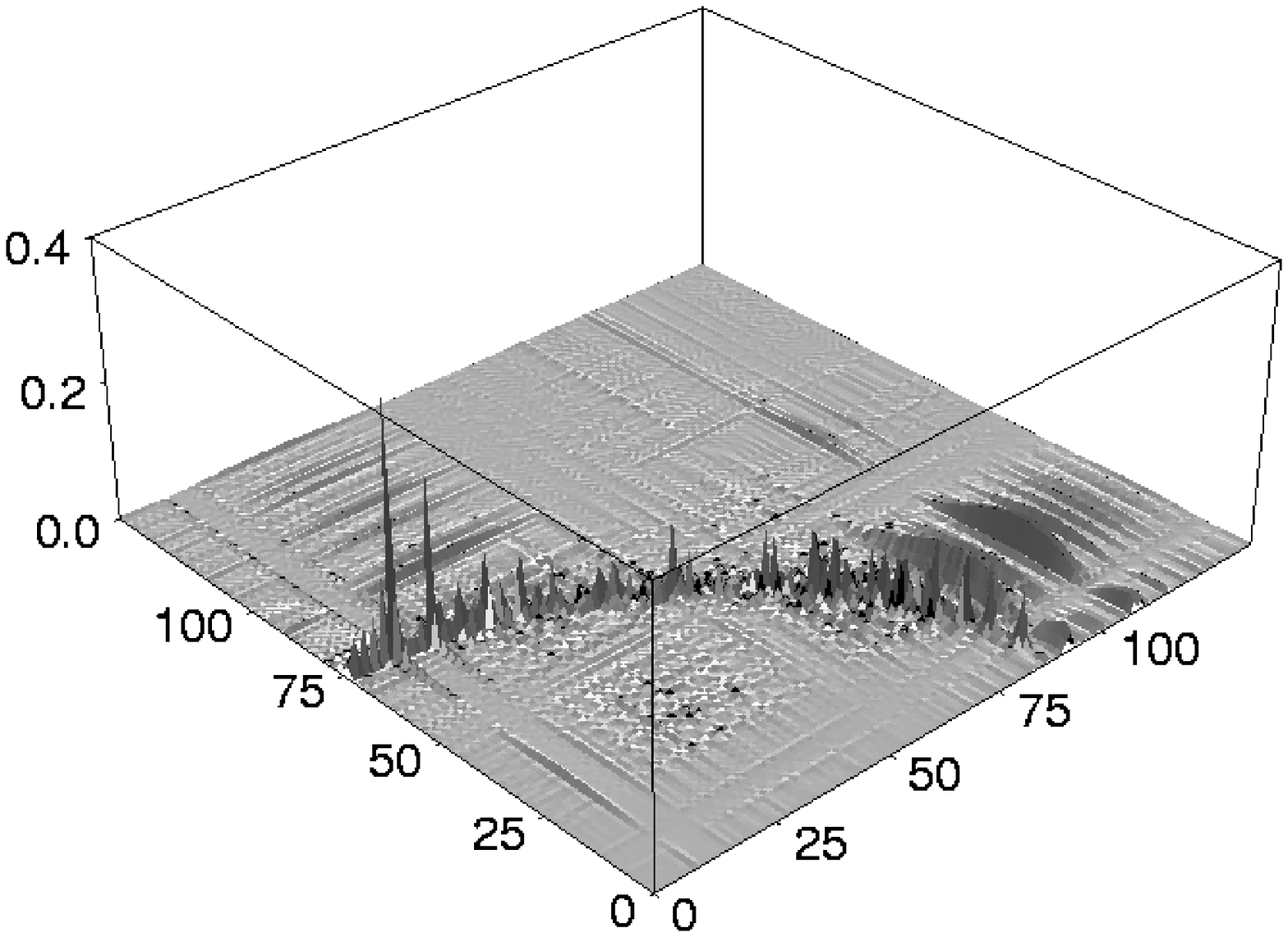}}
\put(4,12){\makebox(2,2){{ \bf $|C^{(IV)}_{n,m}|$}}} 
\put(4,32){\makebox(2,2){{ \bf $|C^{(III)}_{n,m}|$}}} 
\put(4,52){\makebox(2,2){{ \bf $|C^{(II)}_{n,m}|$}}} 
\put(4,72){\makebox(2,2){{ \bf $|C^{(Ia)}_{n,m}|$}}} 
\put(28,2.2){\makebox(2,2){{ \bf $n$}}} 
\put(28,22.2){\makebox(2,2){{ \bf $n$}}} 
\put(28,42.2){\makebox(2,2){{ \bf $n$}}} 
\put(28,62.2){\makebox(2,2){{ \bf $n$}}} 
\put(12.5,2.5){\makebox(2,2){{ \bf $m$}}} 
\put(12.5,22.5){\makebox(2,2){{ \bf $m$}}} 
\put(12.5,42.5){\makebox(2,2){{ \bf $m$}}} 
\put(12.5,62.5){\makebox(2,2){{ \bf $m$}}} 
\put(36,77){\makebox(1,1){{(a)}}}
\put(36,57){\makebox(1,1){{(b)}}}
\put(36,37){\makebox(1,1){{(c)}}}
\put(36,17){\makebox(1,1){{(d)}}}
\end{picture}
\caption[]{\small Structure of the energy surface of the eigenfunctions 
(a) $\Psi^{(Ia)}$, (b) $\Psi^{(II)}$, (c) $\Psi^{(III)}$ and (d) $\Psi^{(IV)}$ 
from Fig.~\ref{fig4}. 
We show the absolute values of the amplitudes $|C^{(\alpha)}_{n,m}|$ on the 
$n,m$-lattice, where $\alpha$ stands for the individual functions labeled $Ia$, 
$II$, $III$ and $IV$. }
\label{fig6}
\end{figure}
\end{center}

\newpage
\unitlength 4.5mm
\vspace*{0mm}
\hspace*{5mm}{
\begin{figure}
\begin{picture}(0,24)
\def\epsfsize#1#2{0.9#1}
\put(2,2){\epsfbox{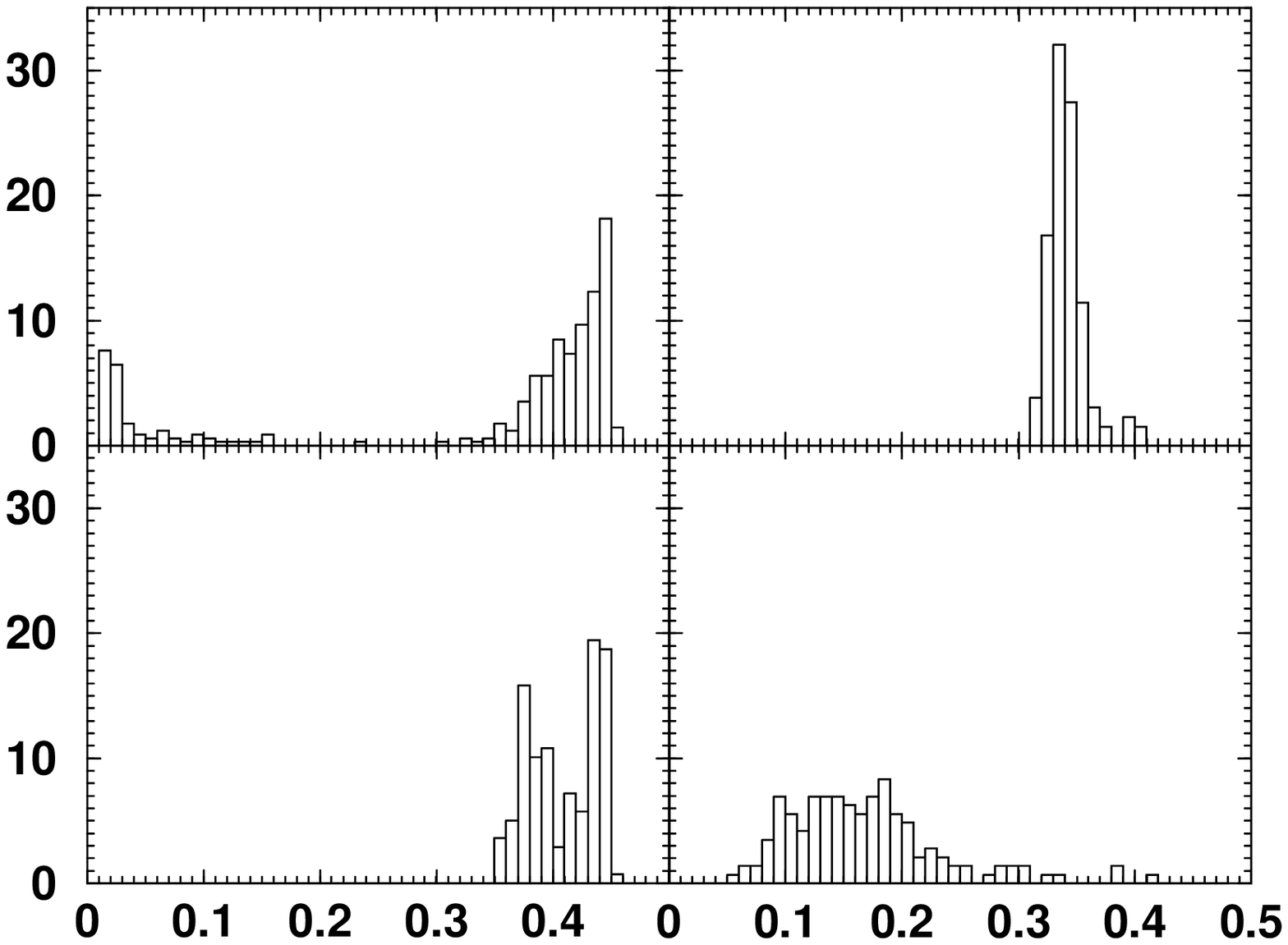}}
\put(12,0){\makebox(1,1){{\Large $V_{\rm{loc}}$}}}
\put(27.5,0){\makebox(1,1){{\Large $V_{\rm{loc}}$}}}
\put(0,22.9){\makebox(1,1){{\Large $p(V_{\rm{loc}})$}}}
\put(0,11.5){\makebox(1,1){{\Large $p(V_{\rm{loc}})$}}}
\put(16,23.5){\makebox(1,1){{\Large \bf (a)}}}
\put(32,23.5){\makebox(1,1){{\Large \bf (b)}}}
\put(16,12){\makebox(1,1){{\Large \bf (c)}}}
\put(32,12){\makebox(1,1){{\Large \bf (d)}}}
\end{picture}
\caption[]{\small The normalized histograms of the $V_{\rm{loc}}$-values
are shown for energy regimes (a) $I$, (b) $II$, (c) $III$ and (d) $IV$. 
For the calculations were used $341$ states in (a), $131$ states in 
(b), $139$ states in (c) and $144$ states in (d).}
\label{fig7}
\end{figure}}

\newpage
\unitlength 3.5mm
\vspace*{0mm}{
\begin{figure}
\begin{picture}(80,20)
\def\epsfsize#1#2{0.8#1}
\put(2,2){\epsfbox{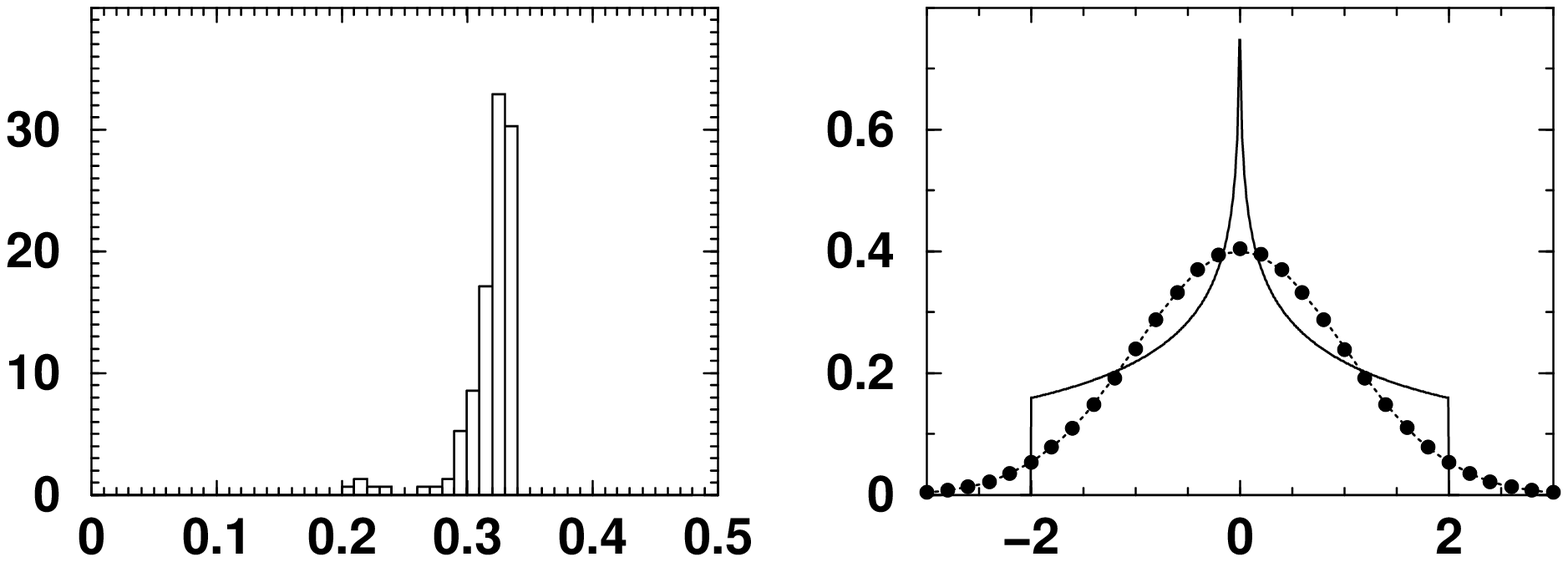}}
\put(13.5,0){\makebox(1,1){{\Large \bf $V_{\rm{loc}}$ }}}
\put(1,16.5){\makebox(1,1){{\Large \bf $p(V_{\rm{loc}})$}}}
\put(35.9,0){\makebox(1,1){{\Large \bf $\psi$}}}
\put(24.5,16.5){\makebox(1,1){{\Large \bf $P(\psi)$}}}
\put(18.5,14.5){\makebox(1,1){{\Large \bf (a)}}}
\put(41.5,14.5){\makebox(1,1){{\Large \bf (b)}}}
\end{picture}
\caption[]{\small (a) The normalized histogram of the $V_{\rm{loc}}$-values is shown for the $152$ eigenstates of the system $R_1$ from the energy interval $[0.225-0.240]$.
(b) The averaged amplitude distribution of the same eigenfunctions as in (a) is indicated by full circles. 
The distributions for a regular $\sin$ or $\cos$-function as well as for
the Gaussian distribution are indicated by a solid and a dotted line,
respecitvely.
}
\label{fig8}
\end{figure}}


\begin{thebibliography} {50}
\bibitem{guhr} T. Guhr, A. M\"uller-Groeling, and H.~A.  
 Weidenm\"uller, Phys. Rep. {\bf 299}, 189 (1998). 
\bibitem{pseudo1} P.~J. Richens and M.~V. Berry, Physica D {\bf 2}, 495 (1981).
\bibitem{pseudo2} B. Eckhardt, J.~Ford and F.~Vivaldi, Physica {\bf 13D}, 339 (1984).
\bibitem{pseudo3} E. Gutkin, Physica D {\bf 19}, 311 (1986).
\bibitem{seba} P.~Seba, Phys. Rev. Lett. {\bf 64}, 1855 (1990). 
\bibitem{cc} T. Cheon and T. D. Cohen, Phys. Rev. Lett.~{\bf 62}, 2769 (1989).
\bibitem{shudoshim} A. Shudo and Y. Shimizu, Phys. Rev. E {\bf 47}, 54 (1993).
\bibitem{shudoetal} A. Shudo, Y. Shimizu, P. Seba, J. Stein, H.-J. St\"ockmann 
and K. Zyczkowski, Phys. Rev. E {\bf 49}, 3748 (1994).
\bibitem{steffi} S. Russ, Phys. Rev. E {\bf 64}, 056240 (2001).
\bibitem{BiswasJain} D. Biswas and S.~R. Jain, Phys. Rev.
A~{\bf42}, 3170 (1990).
\bibitem{BiswasSinha} D. Biswas and S. Sinha, Phys. Rev. E
{\bf 60}, 408 (1999).
\bibitem{shig94} T. Shigehara, Phys. Rev. E {\bf 50}, 4357 (1994).
\bibitem{weaver95} R.~L.~Weaver and D.~Sornette, Phys. Rev. E {\bf 52}, 3341
(1995).
\bibitem{chsh96} T. Cheon and T. Shigehara, Phys. Rev. E {\bf 54}, 3300 (1996).
\bibitem{Borgonovi} F. Borgonovi, G. Casati and B. Li, Phys. Rev. Lett. {\bf 77}, 
4744 (1996).
\bibitem{Frahm1} K.M. Frahm and D.L. Shepelyansky, Phys. Rev. Lett. {\bf 78}, 
1440 (1997).
\bibitem{Frahm2} K.M. Frahm and D.L. Shepelyansky, Phys. Rev. Lett. {\bf 79}, 
1833 (1997).
\bibitem{Lanczos} J. Cullam and R. Willoughby, {\it Lanczos algorithms
for large symmetric eigenvalue computations}, Vol. 1 and 2, Borkh\"auser,
Boston 1985.
\bibitem{kantelhardt98} J.~W. Kantelhardt, A. Bunde,  
 and L. Schweitzer, Phys. Rev. Lett. {\bf 81}, 4907 (1998) and  
 Physica A {\bf 266}, 76 (1999). 
\bibitem{schweitzer99} L. Schweitzer and H. Potempa, Physica A {\bf 266},  486 (1999). 
\bibitem{dyson} F.J. Dyson and M.L. Metha, J. Math. Phys. {\bf 4}, 701 (1963)
and F.J. Dyson and M.L. Metha, J. Math. Phys. {\bf 4}, 713 (1963).
\bibitem{bohigianno1975} O. Bohigas and M.J. Giannoni, Ann. of Phys. {\bf 89}, 
393 (1975).
\bibitem{methabuch} M.~L.~Metha, {\it Random Matrices}, Academic Press, Boston 1991.  
\bibitem{damping} S. Russ and B. Sapoval, Phys. Rev. E {\bf 65}, 036614 (2002). 
\bibitem{biswas} D. Biswas and S. Sinha, Phys. Rev. Lett. {\bf 70}, 916 (1993)
and D.~Biswas, Phys. Rev. E {\bf 54}, R1044 (1996).
\bibitem{Berry} M.V. Berry, J. Phys. A {\bf 10}, 2083 (1977).
\bibitem{McDonald} S.W. McDonald and A.N. Kaufman, Phys. Rev. A {\bf37}, 3067 (1988).
\bibitem{Simmel} F. Simmel and M. Eckert, Physica D {\bf 97}, 517 (1996).
\bibitem{Wegner} F. Wegner, Z. Physik B {\bf 36}, 209 (1980).
\bibitem{Yuri} Y. Hlushchuk, L. Sirko, U. Kuhl, M. Barth and H.-J.~St\"ockmann, 
Phys. Rev.E {\bf 63}, 046208 (2001).
\end{thebibliography}
\end{document}